\documentclass[paper]{JHEP3}
\usepackage{epsfig}
\usepackage{amsmath}
\usepackage{amssymb}
\usepackage{amsbsy}
\usepackage{bbm}
\usepackage{enumerate}
\usepackage{url}

\newlength{\wfig}
\newlength{\hfig}
\newlength{\hfigs}
\def\tr{\mathop{\rm tr}\nolimits}

\def\beq{\begin{equation}}
\def\beqn{\begin{eqnarray}}
\def\eeq{\end{equation}}
\def\eeqn{\end{eqnarray}}

\newcommand\PYTHIA{{\tt PYTHIA}}

\newcommand\HqT{{\tt HqT}}

\newcommand\Phirad{\Phi_{\rm rad}}

\newcommand\muf{\mu_{\sss\rm F}}
\newcommand\mur{\mu_{\sss\rm R}}
\newcommand\mTH{m_{\sss   T}^{{\sss H}}}
\newcommand\mH{m_{\sss  H}}
\newcommand\HThat{\hat{H}_{\sss  T}}
\newcommand\ptrelone{p_{\sss T}^{\sss {\rm rel},j_1}}
\newcommand\ptreltwo{p_{\sss T}^{\sss {\rm rel},j_2}}
\newcommand\ptrel{p_{\sss T}^{\sss {\rm rel},j}}

\def\({\left(} 
\def\){\right)}

\newcommand\sss{\mathchoice%
{\displaystyle}%
{\scriptstyle}%
{\scriptscriptstyle}%
{\scriptscriptstyle}%
}

\newdimen\hbigcirc
\newdimen\wbigcirc

\newdimen\figwidth

\figwidth=\textwidth

\newcommand\as{\alpha_{\sss\rm S}}

\newcommand\pt{p_{\sss T}}
\newcommand\pT{p_{\sss T}}

\newcommand\pzero{p_{\sss\rm 0}}
\newcommand\kt{k_{\sss\rm T}}

\newcommand\POWHEG{{\tt POWHEG}}
\newcommand\POWHEGBOX{{\tt POWHEG BOX}}
\newcommand\MG{{\tt MadGraph4}}
\newcommand\HELAS{{\tt HELAS}}
\newcommand\MCFM{{\tt MCFM}}
\newcommand\MadDipole{{\tt MadDipole}}
\newcommand\MadFKS{{\tt MadFKS}}
\newcommand\MGS{{\tt MadGraphStuff}}
\newcommand\ttilde{\raise.17ex\hbox{$\scriptstyle\mathtt{\sim}$}}

\newcount\minutes 
\newcount\scratch 
 
\def\timestamp{%
\scratch=\time 
\divide\scratch by 60 
\edef\hours{\the\scratch} 
\multiply\scratch by 60 
\minutes=\time 
\advance\minutes by -\scratch 
---$\,$\hours:\null 
\ifnum\minutes< 10 0\fi 
\the\minutes}

\preprint{FERMILAB--PUB--12--040--T\\
          {CERN--PH--–TH--2012--–048}\\
          {ZU--TH 03/12}}

\title{NLO Higgs boson production plus one and two jets
using the POWHEG BOX, MadGraph4 and MCFM}

\author{John M. Campbell\\
Fermilab, Batavia, IL 60510, USA\\
  E-mail: \email{johnmc@fnal.gov}}

\author{R. Keith Ellis\\
Fermilab, Batavia, IL 60510, USA\\
  E-mail: \email{ellis@fnal.gov}}

\author{Rikkert Frederix\\
  Institut f\"ur Theoretische Physik,
  Universit\"at Z\"urich, Winterthurerstrasse 190, CH-8057 Z\"urich,
  Switzerland\\
  E-mail: \email{frederix@physik.uzh.ch}}

\author{Paolo Nason\\
  INFN, Sezione di Milano-Bicocca,
  Piazza della Scienza 3, 20126 Milan, Italy\\
  E-mail: \email{Paolo.Nason@mib.infn.it}}

\author{Carlo Oleari\\
  Universit\`a di Milano-Bicocca and INFN, Sezione di Milano-Bicocca\\
  Piazza della Scienza 3, 20126 Milan, Italy\\
  E-mail: \email{Carlo.Oleari@mib.infn.it}}

\author{Ciaran Williams\\
Fermilab, Batavia, IL 60510, USA\\
  E-mail: \email{ciaran@fnal.gov}}

\abstract{ We present a next-to-leading order calculation of Higgs boson
  production plus one and two jets via gluon fusion interfaced to shower
  Monte Carlo programs, implemented according to the \POWHEG{} method.  For
  this implementation we have used a new interface of the \POWHEGBOX{} with
  \MG{}, that generates the codes for generic Born and real processes
  automatically.  The virtual corrections have been taken from the \MCFM{}
  code.
  We carry out a simple phenomenological study of our generators, comparing
  them among each other and with fixed next-to-leading order results.

}
\keywords{QCD, Monte Carlo, NLO Computations, Resummation, Collider Physics
\vfill 
\vfill 
}

\begin{document}

\section{Introduction}
The search for the Standard Model Higgs boson is entering the endgame phase.
Since the start of the LHC running, the allowed mass range for the Standard
Model Higgs boson has been greatly curtailed. Furthermore there are
tantalizing, but inconclusive, hints in the remaining low mass
region~\cite{Collaboration:2012si, Chatrchyan:2012tx}. Should these hints be
confirmed by further data, it will be a matter of some urgency to examine the
properties of the new state (or states). In the low mass region, the Standard
Model Higgs boson is predicted to have about ten decay modes with
branching fractions greater than one per mille, so there will be a number of
channels to be studied.

In order to extract the Higgs boson couplings to fermions and gauge bosons
from these different channels,
detailed information about the production rates will be
required~\cite{Zeppenfeld:2000td, Plehn:2001nj, Duhrssen:2004cv}.  In this
context, the Higgs boson + 2 jet process enters principally as an irreducible
background for the vector boson fusion~(VBF) process~\cite{Figy:2003nv,
  Ravindran:2002dc, Berger:2004pca}. This process is of particular importance
in the extraction of the $HWW$ and $HZZ$ couplings and in confirming if the
detected scalar particle is the Higgs boson responsible for electroweak
symmetry breaking.  In order to distinguish VBF Higgs boson signal from
backgrounds, stringent cuts are required on the Higgs boson decay products as
well as on the two forward quark jets which are characteristic for VBF.
Information on the CP properties of the Higgs boson can be extracted by
studying the azimuthal distributions of the two hardest
jets~\cite{Plehn:2001nj, Klamke:2007cu}, in events with a large ($\gtrsim 3$
units) rapidity separation between them.  In fact, the signature of the VBF
Higgs boson production are two jets well separated in rapidity. In addition,
we expect very low hadronic activity between the two hardest jets, due to the
exchange of colourless vector bosons in the $t$ channel, in contrast to what
is expected for the gluon-fusion production mechanism.  For the gluon-fusion
processes, the higher-order corrections are significant~\cite{DelDuca:2004wt,
  DelDuca:2006hk, Campbell:2006xx, Campbell:2010cz}, and a detailed
understanding of the structure of the radiation pattern is needed in order to
access the efficiency of the central-jet veto~\cite{DelDuca:2001eu,
  DelDuca:2001fn}.


The Higgs boson search channels are often subdivided according to the number
of associated jets, for a number of reasons.  First, a differing number of
associated jets can imply a different source of background. For example, in
the search for a Higgs boson decaying to $W^+ W^-$, the backgrounds have
different origins in the differing jet bins, so it proves advantageous to
analyze the different jet multiplicities separately. Second, in the two-jet
bin, the new VBF production channel will come into play. For
these reasons it is pressing to provide NLO predictions for gluon-fusion
initiated Higgs boson + jets processes in
the \POWHEG{} formalism, so that experimentalists can incorporate
the best theoretical information into their analyses.

Phenomenological studies for the production of a Higgs boson in association
with two jets, at the parton level and at fixed order, have been available
for a while.  In refs.~\cite{DelDuca:2001eu, DelDuca:2001fn}, the calculation
was performed at leading order, with the gluons coupling to the Higgs boson
via a top-quark loop, retaining the exact top-quark mass  dependence ($m_t$)
in the whole calculation. In that paper it was shown that the large-$m_t$
limit provides an excellent approximation to the full $m_t$ dependence when
the Higgs boson mass, $\mH$, is small compared to the top-pair threshold.
The large-$m_t$ limit was found to break down for $\mH>m_t$ and when jet
transverse momenta become large ($\pT^{\sss j}\gtrsim m_t$). However, large dijet
invariant masses do not invalidate the $m_t\to\infty$ limit, as long as
the Higgs boson mass and the jet transverse momenta are small enough, less
than the top-quark mass in practice.  
This observation opened the possibility to compute the NLO corrections to
Higgs boson + 2~jet production in gluon fusion, in the large-$m_t$ limit,
starting with an effective coupling of the Higgs boson to the gluon
field~\cite{Campbell:2006xx, Campbell:2010cz}

In this work we have implemented Higgs boson plus one- and two-jet production
in gluon fusion, in the large top-quark mass limit, using the \POWHEG{}
method. While for the first process a matched NLO+shower calculation has
already appeared in the literature~\cite{Hoeche:2011fd}, the second one has
never been performed before.  We have built our generators using the
\POWHEGBOX{} framework~\cite{Alioli:2010xd}, with the virtual corrections
taken from the \MCFM{} program~\cite{MCFM}, and the Born , colour-correlated
Born, spin-correlated Born and real contributions computed using a new
\POWHEGBOX{} interface to the \MG{}~\cite{Stelzer:1994ta, Alwall:2007st}
program, which is fully generic and can be applied to any process. Thus, the
aim of this paper is twofold:
\begin{itemize}
\item To present a new generic interface of \MG{} to the \POWHEGBOX{} that
  allows for the automatic generation of the code for the Born amplitude, the
  Born colour- and spin-correlated contributions, for the real-radiation
  amplitude and for the Born colour flow in the leading-colour
  approximation. These ingredients are all needed in the implementation of a
  new process in the \POWHEGBOX{}. This interface can be used for any process
  that can be generated with \MG{}.  Thanks to this interface, in order to
  construct a \POWHEGBOX{} generator, one needs only to provide the Born
  phase space and the virtual corrections.
\item  To illustrate the Higgs plus one and two jet generators, by comparing
  their outputs to the corresponding NLO results and amongst themselves.
\end{itemize}
The paper is organized as follows: in sec.~\ref{sec:MGinterface}, we present
the new interface of the \POWHEGBOX{} to \MG{}. In sec.~\ref{sec:ggH}, we
discuss the approximations used in our calculation and how we extract the
virtual corrections from the \MCFM{} code.  In sec.~\ref{sec:HJ_HJJ} we give
more details about the phase-space generator for Higgs boson plus one and two
jets, and how we deal with the divergences present at the Born level. After 
briefly recalling a few crucial \POWHEGBOX{} parameter settings, we discuss
some phenomenology in sec.~\ref{sec:phenomenology}. In particular we compare the NLO
differential cross sections with the results obtained with the \POWHEGBOX{},
at several levels of approximation.  Finally, in sec.~\ref{sec:conclusions},
we summarize our findings. A few technical details of the \MG{} interface and
of the \PYTHIA{} setup are collected in the two appendices.

\section{The interface to \MG}
\label{sec:MGinterface}
The \MG\ package~\cite{Stelzer:1994ta, Alwall:2007st} generates squared
tree-level matrix elements in an automatic way. It is based on Feynman
diagrams, using \HELAS{} routines~\cite{Murayama:1992gi} as building blocks
which are subsequently combined into colour-ordered amplitudes and written
in a FORTRAN code. Furthermore, the \MadDipole~\cite{Frederix:2008hu,
  Frederix:2010cj, Gehrmann:2010ry} extension of \MG\ and the
\MadFKS\ tool~\cite{Frederix:2009yq} can generate the spin- and
colour-correlated Born squared amplitudes. The \MG{} package can thus
construct all the ingredients needed to interface a NLO calculation to the
\POWHEGBOX{} (see the complete list in the introduction of
ref.~\cite{Alioli:2010xd}), with the exception of the Born phase space and
the virtual matrix elements.  This makes the \MG\ framework an ideal
environment to complement the \POWHEGBOX\ in the creation of all tree-level
squared matrix elements and in this section we describe a newly developed
interface between these two codes.

For a given process, the routines and parameters that are generated by
\MG\ for the \POWHEGBOX\ are (see ref.~\cite{Alioli:2010xd} for more
details on the notation and conventions used):
\begin{itemize} 
\item the multiplicity of the Born and real-emission processes,
  \verb|nlegborn| and \verb|nlegreal|, respectively, defined in the
  \verb|nlegborn.h| include file;

\item the list of Born and real-emission flavour structures\\
  \verb|flst_born(k=1,nlegborn,j=1:flst_nborn)|,   \\
  \verb|flst_real(k=1,nlegreal,j=1:flst_nreal)|,   \\
  respectively, to be initialised by a call to the \verb|init_processes|
  routine;

\item the routine\\
  \verb|setborn(p(0:3,1:nlegborn),bflav(1:nlegborn),born,|\\
  \verb|        bornjk(1:nlegborn,1:nlegborn),bmunu(0:3,0:3,1:nlegborn))|\\
   which computes, for a given set of four-momenta \verb|p| and flavour
   structure \verb|bflav|, the Born squared matrix element \verb|born|, the
   colour-correlated, \verb|bornjk|, and the spin-correlated one,
   \verb|bmunu|;

\item the routine that computes the real emission squared matrix elements
  \verb|amp2| (stripped of a factor $\as/(2\pi)$), for a given momentum
  configuration \verb|p| and flavour structure \verb|rflav|\\
  \verb|setreal(p(0:3,nlegreal),rflav(1:nlegreal),amp2)|;

\item a colour flow assignment to the Born squared matrix elements, in the
  leading colour approximation, needed by the showering programs. The colour is
  assigned on statistical grounds, based on information that is
  cached during the call to the squared matrix elements, by the routine
  \verb|borncolour_lh|. In addition to this routine, the interface also provides
  a similar one for the real squared matrix elements,
  \verb|realcolour_lh|. In the present version of the \POWHEGBOX{} this routine is not used;
  instead a different method is employed to assign the colour to the real-radiation matrix
  elements, as discussed in sec.~8 of ref.~\cite{Alioli:2010xd};

\item an interface to the Les Houches parameter input card to specify the
  physics-model parameters through the routine \verb|init_couplings|.
\end{itemize}
In contrast to the default way of generating amplitudes with \MG,
all the squared matrix elements for the various flavour structures are
written in the same directory, making sure that the files and routines
have different names. This allows one to compile the code into a
single library that contains all the matrix elements for all the
flavour structures. An interface is provided that concatenates the
flavour vectors \verb|bflav| or \verb|rflav| into strings, which are
used as unique identifiers of each of the squared matrix elements.

More technical details on the use of the \MG{} interface to the \POWHEGBOX{}
can be found in appendix~\ref{app:tech_details}.

\section{Higgs boson production in gluon fusion}
\label{sec:ggH}
The amplitudes for a Higgs boson in association with three,
four or five partons, needed for the Higgs boson plus one and two jet cross
sections at NLO, are calculated using an effective Lagrangian to express the
coupling of the gluons to the Higgs field~\cite{Wilczek:1977zn}
\begin{equation} 
\label{EffLag}
\mathcal{L}_{\sss H} = \frac{C}{2} \, H\,\tr
G_{\mu\nu}\,G^{\mu\nu}\,,
\end{equation}
where the trace is over the colour degrees of freedom.  At the order required
in this paper, the coefficient $C$ is given in the $\overline{\rm MS}$ scheme
by~\cite{Djouadi:1991tka,Dawson:1990zj}
\begin{equation}
C =\frac{\as}{6 \pi v} \( 1 +\frac{11}{4 \pi} \as\)
 + {\cal O}(\as^3) \,,
\end{equation}
where $v$ is the vacuum expectation value of the Higgs field, $v = 246$ GeV.


We have used the automatic interface described in sec.~\ref{sec:MGinterface}
to generate the code for the Born amplitude, the Born colour- and
spin-correlated contributions and for the real cross section. In
appendix~\ref{app:tech_details} we give some more details on the adopted
procedure. The one-loop amplitudes for the Higgs boson plus three and four
parton processes are extracted from \MCFM{} as described below.

\subsection{Virtual cross sections}
\label{sec:virtual}
The complete set of one-loop amplitudes for all Higgs boson + 4 parton processes
have been available~\cite{Berger:2006sh, Badger:2006us,
  Badger:2007si,Glover:2008ffa, Badger:2009hw, Dixon:2009uk, Badger:2009vh} for some time.
These formulae have been implemented into \MCFM~\cite{Campbell:2010cz}.
  
The interface between \MCFM{} and the \POWHEGBOX{} is fairly straightforward.
Since in the \POWHEGBOX{} implementation the Born, real and subtraction terms
are available independently, \MCFM{} needs only to return the pure virtual
contributions.  The interface to \MCFM{} transfers the electroweak
parameters, scale and scheme choices.  Once these are established, calls to
the virtual routines of \MCFM{} can be made on an event-by-event basis.
  
Since the routines that fill the values of the electroweak parameters (and
the other information) are generic, this interface could be fairly easily
extended to include other processes currently implemented in \MCFM. However
the normal \MCFM{} routines are designed to return matrix elements that are
summed over the flavour of the final-state partons. Therefore, in order to
correctly interface to the \POWHEGBOX{}, one must make small modifications to
the \MCFM{} code such that it returns the matrix elements for individual
final-state flavour combinations.

\section{Implementation of the $Hj$ and $Hjj$ generators}
\label{sec:HJ_HJJ}

In the following, we use the notation $H$, $Hj$ and $Hjj$ to refer to Higgs
boson generators that, at the Born level, describe the production of a Higgs
boson plus zero, one and two partons.

Using the automated \MG{} interface described in sec.~\ref{sec:MGinterface},
we have built the routines for the Born amplitude, the Born colour- and
spin-correlated contributions and the real squared amplitudes, directly in
the format required by the \POWHEGBOX{}. The generation of the amplitudes is
fast, taking only a few minutes for the $Hjj$ process. The virtual
corrections were extracted from the \MCFM{} code, as described in
sec.~\ref{sec:virtual}. At this point, the only missing ingredient for
completing our generators is the Born phase space.

\subsection{Phase space for the $Hj$ generator}
The phase space routine for the $Hj$ Born process is trivial, since it is
just a two-body phase space. It has been implemented with the possibility
to activate an optional cut on the transverse momentum of the Higgs boson, and
with an effective importance sampling of the small transverse-momentum region.
The cut is necessary if one wants to generated unweighted events. Alternatively
the cut can be set to zero, and the cross section is weighted with a suppression
factor,  that is a function of the underlying-Born kinematics. This factor is equal to
\begin{equation}
\label{eq:F_funct}
F=\frac{\pt^2}{\pzero^2+\pt^2}\,,
\end{equation}
where $\pt$ is the Higgs boson transverse momentum in the underlying-Born
configuration and $\pzero$ is set by the user~\footnote{The value of
  $\pzero$, set to 20~GeV for the simulation done in this paper, or the form
  of the suppression factor $F$ in eq.~(\ref{eq:F_funct}), can be changed by
  the user by modifying the {\tt born\_suppression} routine in the {\tt
    Born\_phsp.f} file.}. Events are then generated uniformly in the cross
section times $F$, and with a weight proportional to $1/F$. The normalization
of the weight is such that the cross section for events passing a set of cuts
is given by the sum of all weights for the events passing the cut divided by
the total number of generated events.

\subsection{Phase space for the $Hjj$ generator}
The phase space for the $Hjj$ generator was built using the same factorized
phase space that \POWHEG{} uses for the real kinematics.  In other words, we
treat the $Hjj$ Born phase space as the real phase space for the $Hj$ process
with one extra emission.  We have allowed the possibility of
performing importance sampling in the regions where the Born cross section
becomes singular. In more detail, labelling 1 and 2 as the two final-state
light partons in the $Hjj$ process, we write the $Hjj$ phase space using the
following identity
\begin{equation}
 d\Phi_{H12}=d\Phi_{H12}\, \frac{N}{d_{12}}\, \frac{E_1}{E_1+E_2}
+d\Phi_{H12} \, \frac{N}{d_{12}}\,\frac{E_2}{E_1+E_2}
+d\Phi_{H12} \, \frac{N}{d_{1}}
+d\Phi_{H12} \, \frac{N}{d_{2}}\,, 
\label{eq:phspspit}
\end{equation}
where $E_i$ is the energy of the $i$-th parton in the center-of-mass frame, and
\begin{equation}
N=\left(\frac{1}{d_{12}}+\frac{1}{d_{1}}+\frac{1}{d_{2}}\right)^{-1},
\end{equation}
where $d_1$, $d_2$ and $d_{12}$ are phase-space functions that vanish
respectively when parton 1 is collinear to the initial-state partons, parton
2 is collinear to the initial-state partons, and partons 1 and 2 are
collinear to each other. Their precise form is given in eqs.~(4.23)-(4.26) of
ref.~\cite{Alioli:2010xd}.  We then factorize each phase-space factor
in~(\ref{eq:phspspit}) according to the formula
\begin{eqnarray}
 d\Phi_{H12}&=&d\Phi_{H2}\,d\Phirad^{(21)} \,\frac{N}{d_{12}}\,\frac{E_1}{E_1+E_2}
+d\Phi_{H1}\, d\Phirad^{(12)} \, \frac{N}{d_{12}}\, \frac{E_2}{E_1+E_2}
\nonumber \\
&+& d\Phi_{H2}\, d\Phirad^{(01)} \, \frac{N}{d_{1}}
+d\Phi_{H1}\, d\Phirad^{(02)}\, \frac{N}{d_{2}}\,.
\label{eq:phspspit1}
\end{eqnarray}
The subscripts $H1$ or $H2$ characterize the underlying Born, while the
superscript in $\Phirad$ specifies the radiation process.  Thus, for example,
$d\Phi_{H2}$ is the underlying-Born phase space of the Higgs boson together
with parton 2, and $d\Phirad^{(21)}$ is the radiation phase space
corresponding to parton 2 emitting parton 1. The notation $d\Phirad^{(01)}$
or $d\Phirad^{(02)}$ means that partons 1 or 2 are emitted by the
initial-state partons. The factorization of the phase space into underlying
Born and radiation phase space for both initial- and final-state radiation is
the default one used in \POWHEG{}, and is described in detail in secs.~5.1
and~5.2 of ref.~\cite{Frixione:2007vw}. The decomposition in
eq.~(\ref{eq:phspspit1}) is such that appropriate importance sampling is
performed in all singular regions. In fact, the factors in each term damp all
but one singular region, and the corresponding factorized phase space
performs importance sampling precisely in that region. Ideally, the
phase-space integration should be performed with an integrator that can sum
over a discrete variable. The \POWHEGBOX{} integrator~\cite{Nason:2007vt}
does not have this feature at the moment.  Thus, we divided the range of one
extra integration variable into four segments, mapping each segment to one of
the phase space components.

The phase space of eq.~(\ref{eq:phspspit1}) is not the default one
used in the $Hjj$ generator. It is activated by setting the variable
{\tt fullphsp} in the {\tt powheg\_input} file. The default phase
space is simply
\begin{equation}
d\Phi_{H12}=d\Phi_{H1} \, d\Phirad^{(02)}\,,
\end{equation}
with no importance sampling at all.  We have in fact observed that the loss
of efficiency due to the increased number of calls to the matrix-element
routines overwhelms any benefit arising from the improved importance
sampling.

As for the case of the $Hj$ generator, the $Hjj$ generator also includes the
implementation of a Born suppression factor for the suppression of the
singularities of the underlying-Born amplitude, in the {\tt
  born\_suppression} routine. It has the form
\begin{equation}
F=\( \frac{1/\pzero^2}{1/p_{\sss T1}^2+1/p_{\sss T2}^2+1/p_{\sss T12}^2+ 1/\pzero^2}\)^2,
\end{equation}
where $\pzero$ is a parameter that characterizes the minimum jet energy where
some accuracy is required, $p_{\sss T1}$ and $p_{\sss T2}$ are the transverse momenta
(with respect to the beam axis) of the two final-state partons, and
\begin{equation}
p_{\sss T12}^2=2\, \(1-\cos\theta_{12}\) \,\frac{E_1^2\,E_2^2}{E_1^2+E_2^2}\,,
\end{equation}
that can be interpreted as the transverse momentum of parton 1 with respect
to parton 2 or vice-versa, depending upon which is the softest.  This
suppression factor can also play the role of a generation cut, if $\pzero$ is
chosen small enough.

\subsection{The \POWHEGBOX{} parameter setting}
We have turned on the {\tt bornzerodamp} flag by default in both the $Hj$ and
the $Hjj$ generators. The purpose of this flag is explained in
ref.~\cite{Alioli:2010xd}.  This results in a considerable speedup of the
$Hjj$ code.  However, no appreciable differences were observed in the
results obtained without the {\tt bornzerodamp} option.

The separation of the singular regions is controlled in the \POWHEGBOX{} by
the parameters {\tt par\_diexp} and {\tt par\_dijexp} that are set to $1$ by
default~(see ref.~\cite{Alioli:2010xd}, section~4.7).  For the results
presented here we have chosen to set them to $2$, which leads to slightly
better stability for the set of distributions that we have considered. We
note that these parameters were also set to $2$ in the \POWHEGBOX{} dijet
generator~\cite{Alioli:2010xa}. Higher values of these parameters correspond
to sharper separation of singular regions.

\section{Phenomenology}
\label{sec:phenomenology}
In this section, we present a phenomenological study of our generators. This
study does not aim to produce results with realistic experimental cuts, but
rather to compare the predictions of the generators for some key
observables. In particular, we compare the generator results to the fixed NLO
result and with the \POWHEG{} output at the level of the generation of the
hardest radiation (i.e.~before interfacing the result to a parton shower
program), after the shower with no hadronization effects, and after
hadronization.

We present results for the LHC running at 7~TeV, computed using the CTEQ6M
parton distribution function~(pdf) set~\cite{Pumplin:2002vw}.  The same
calculations can easily be performed with any other available
set~\cite{Martin:2009iq, Ball:2008by}, but a study of pdf effects is beyond
the scope of the present paper.  Jets are reconstructed using the anti-$\kt$
jet algorithm~\cite{Cacciari:2008gp}, with $R=0.5$ and default recombination
scheme. Cuts of 20, 50 and 100~GeV on the final-state jets are considered.

The factorization- and renormalization-scale choice deserves a more detailed
discussion.  In view of the large NLO corrections for these processes,
the scale dependence is quite large. This is also a consequence of the fact
that Higgs boson production in gluon fusion starts at order $\as^2$, so that
the $Hj$ and the $Hjj$ processes are of order $\as^3$ and $\as^4$,
respectively. While for $H$ production the natural scale choice is of the
order of the Higgs boson mass, in the case of $Hj$ production one may have a
two-scale problem, if the transverse momentum of the jet is much smaller or
much larger than the Higgs boson mass. In addition, besides the Higgs boson
mass, one may also consider the Higgs boson transverse momentum, which may be
more appropriate for very small or very large Higgs boson transverse momentum
$\pt^{\sss H}$, or the Higgs boson transverse mass
$\mTH=\sqrt{\mH^2+(\pt^{\sss H})^2}$, which may be more appropriate for large
Higgs boson transverse momentum. For the $Hjj$ case, there are even more
possibilities. Although a full study of scale dependence would be very
valuable, we will not perform it in the present work.  In the $Hj$ case we
will limit ourselves to the fixed Higgs boson mass scale choice and to
transverse momentum of the Higgs boson in the underlying-Born
configuration. In the $Hjj$ case, we will consider the fixed Higgs boson mass
scale choice and to the $\HThat$ scale choice at the underlying-Born level,
where
\begin{equation}
\HThat=\mTH+\sum_i {\pt}_i
\end{equation}
and ${\pt}_i$ are the final-state parton transverse momenta in the
underlying-Born kinematics.  All results shown in the next sections have been
computed with $\mH=120$~GeV and $\Gamma_{\sss H}= 0.00575$~GeV. The Higgs
boson momentum has been generated distributed according to a Breit-Wigner
function, with fixed width.

In the following, we will label ``LHE'' the results obtained at the level of the
\POWHEG{} first emission, ``PY'' the results obtained with the
\POWHEG{}+\PYTHIA{} combination with the hadronization and underlying event
switched off, and with ``NLO'' the fixed NLO results.

For some observables, there is overlap among the various generators. For
example, the Higgs boson transverse momentum, as well as the one-jet
multiplicity and the transverse momentum of the leading jet, are described by
both the $H$ and the $Hj$ generators.  The $H$ generator gives a plausible
description for these quantities also at small transverse momenta, since it
provides a resummation of transverse-momentum logarithms that the $Hj$
program does not provide.  On the other hand, at large transverse momenta,
the $Hj$ generator has NLO accuracy, something that the $H$ generator does
not have.  For the same quantities, the $Hjj$ generator cannot be used, since
it requires the presence of a second jet. The two-jet multiplicity, as well
as the second hardest jet transverse-momentum distribution, are provided by
both the $Hj$ and the $Hjj$ programs, again with a different level of
accuracy.  A comparison of the generators in the regions where they overlap
will be carried out in sec.~\ref{sec:comparison}.

\subsection{Results for $Hj$ production}
We have generated a sample with 2M events at the fixed scale and 7.5M at the
running scale.  The event generation time is approximately 25 seconds for
1000 events on a typical CPU.

We have simulated $Hj$ production with two scale choices: $\muf=\mur=\mH$ and
the $\pt$ of the underlying-Born parton $\muf=\mur=\pt^{\sss\rm UB}$.  All
the following plots will come in pairs, with the left plot referring to
$\muf=\mur=\mH$, and the right one to $\muf=\mur=\pt^{\sss\rm UB}$.  We
compare the fixed NLO results, the \POWHEG{} hardest-emission results~(LHE)
and \POWHEG{}+\PYTHIA{}~(PY) ones, where the hadronization and underlying-event effects
in \PYTHIA{} have been turned off (see appendix~\ref{app:PY_setup}). We show results for a
few physical observables with three different cuts applied to
the final-state jets: $\pt$ cuts of $20$, $50$ and $100$~GeV.

\begin{figure}[htb]
\begin{center}
\includegraphics[height=\hfig]{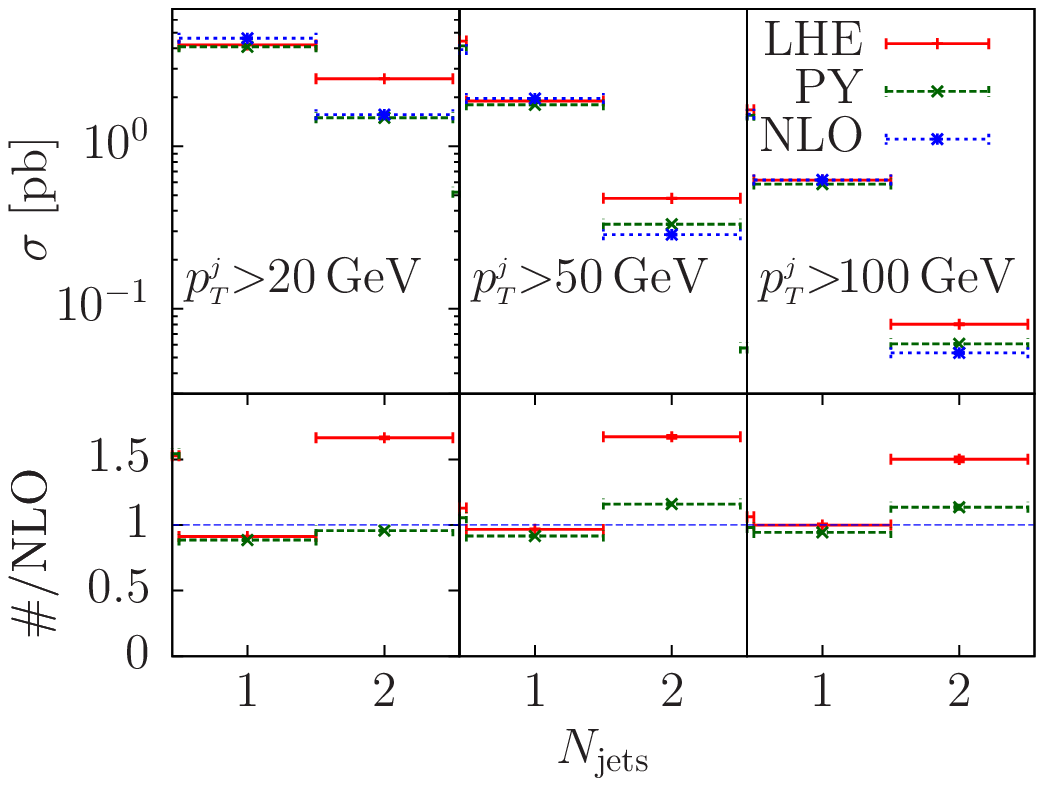} 
\includegraphics[height=\hfig]{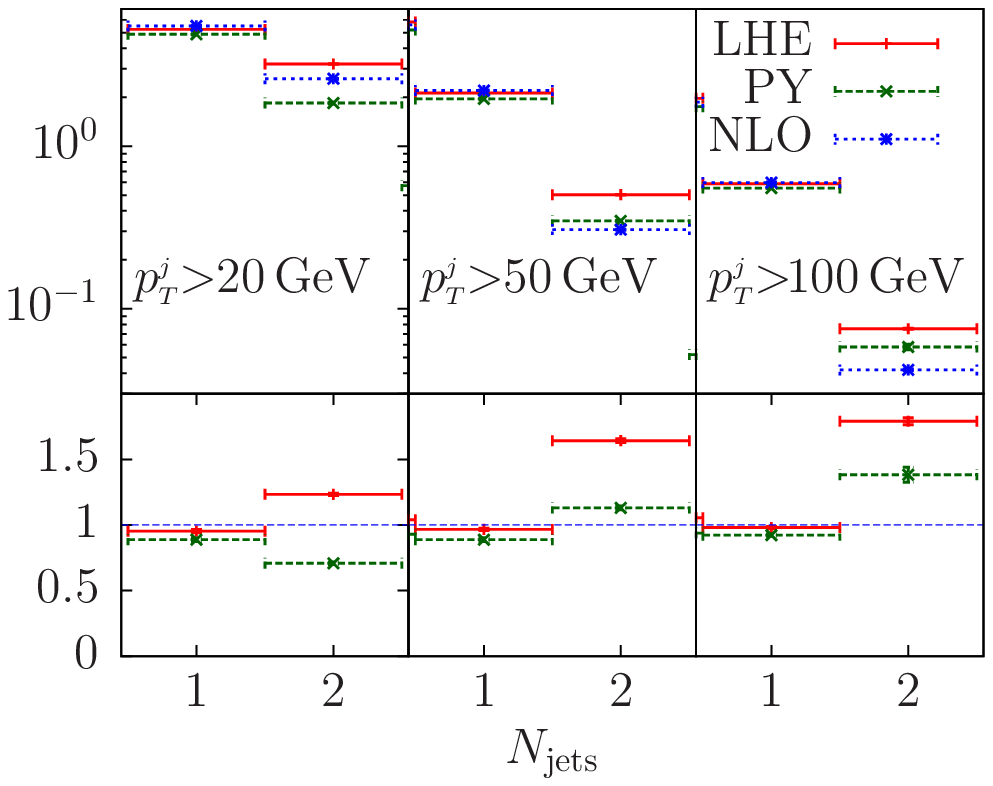} 
\caption{One- and two-jet multiplicities for the \POWHEG{}
  hardest-emission~(LHE) results, the \POWHEG{} events showered with
  \PYTHIA{}~(PY) and the fixed NLO result~(NLO).  The left plot is computed
  using $\muf=\mur=\mH$, while in the right plot $\muf=\mur=\pt^{\sss\rm
    UB}$.}
\label{fig:HJ_mult}
\end{center}
\end{figure}
In fig.~\ref{fig:HJ_mult} we compare the jet multiplicity for one and
two jets, at the NLO, LHE and PY levels, for jet cuts of 20, 50 and 100~GeV. We
notice that the one-jet multiplicity is similar in the three results. 
The \PYTHIA{} multiplicity is smaller than the LHE, which can be understood as a result of showering
off the first jet.  More marked differences can be seen in the cross
sections with two jets: here the LHE result is clearly larger than the
NLO one, showing however a very different pattern as a function of the jet
cut, depending upon the scale choice. We will clarify the origin of these
patterns later, when we discuss the transverse-momentum spectrum of the jets.
We point out, however, that the NLO result for the two-jet multiplicity
is of order $\as^4$, and thus has a marked scale dependence.
In the left plot the scale is equal to the Higgs boson mass, whilst in the plot on the right 
the scale is of the order of the minimum jet $\pT$. For a jet cut of 20~GeV these scales 
are widely separated, which explains the large difference in the NLO results.

\begin{figure}[htb]
\begin{center}
\includegraphics[height=\hfigs]{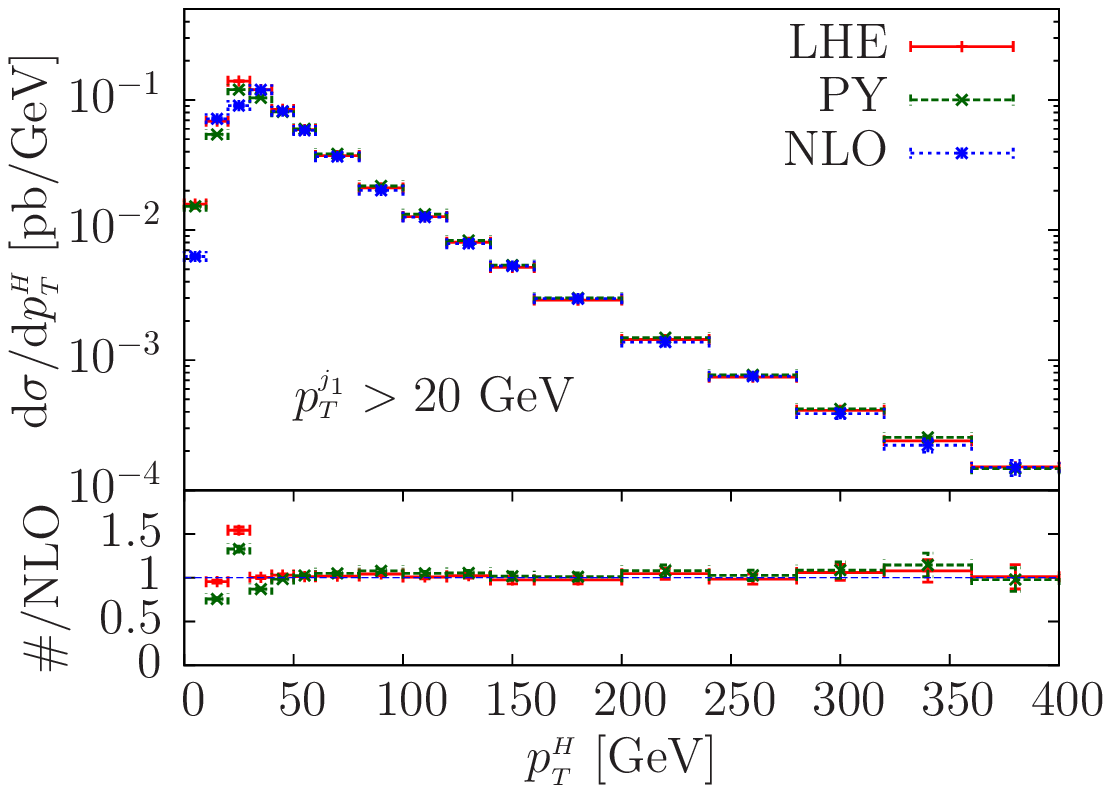}
\includegraphics[height=\hfigs]{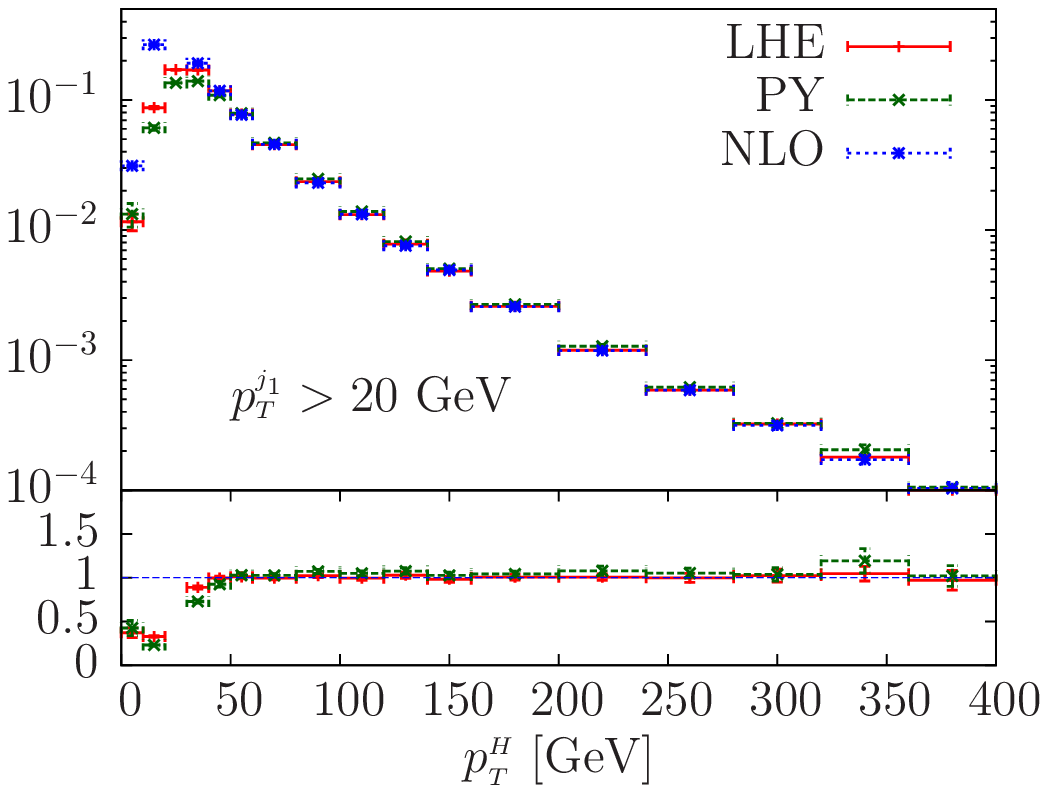} 
\caption{Transverse momentum of the Higgs boson.  The notation is as in
  fig.~\ref{fig:HJ_mult}.  }
\label{fig:HJ_h-pt}
\end{center}
\end{figure}
In fig.~\ref{fig:HJ_h-pt} we show the transverse momentum of the Higgs boson.
The LHE, the PY and the NLO results are in remarkable
agreement for this quantity, which is expected, due its inclusiveness.  
\begin{figure}[htb]
\begin{center}
\includegraphics[height=\hfigs]{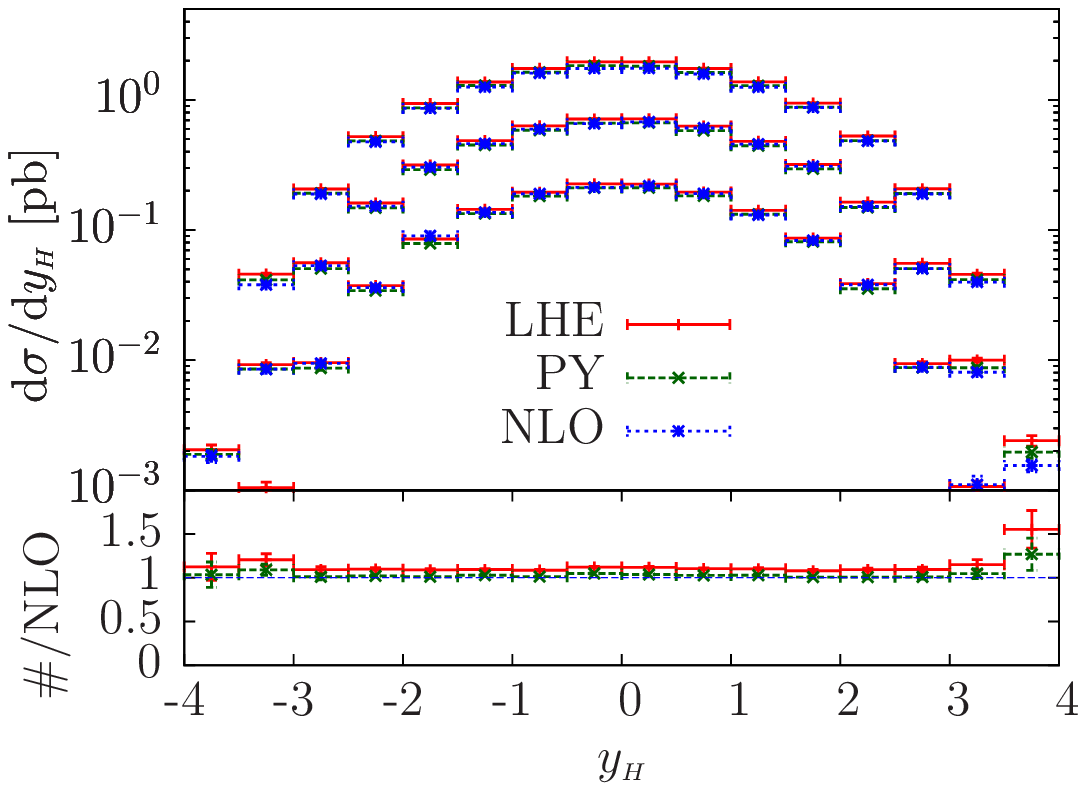}
\includegraphics[height=\hfigs]{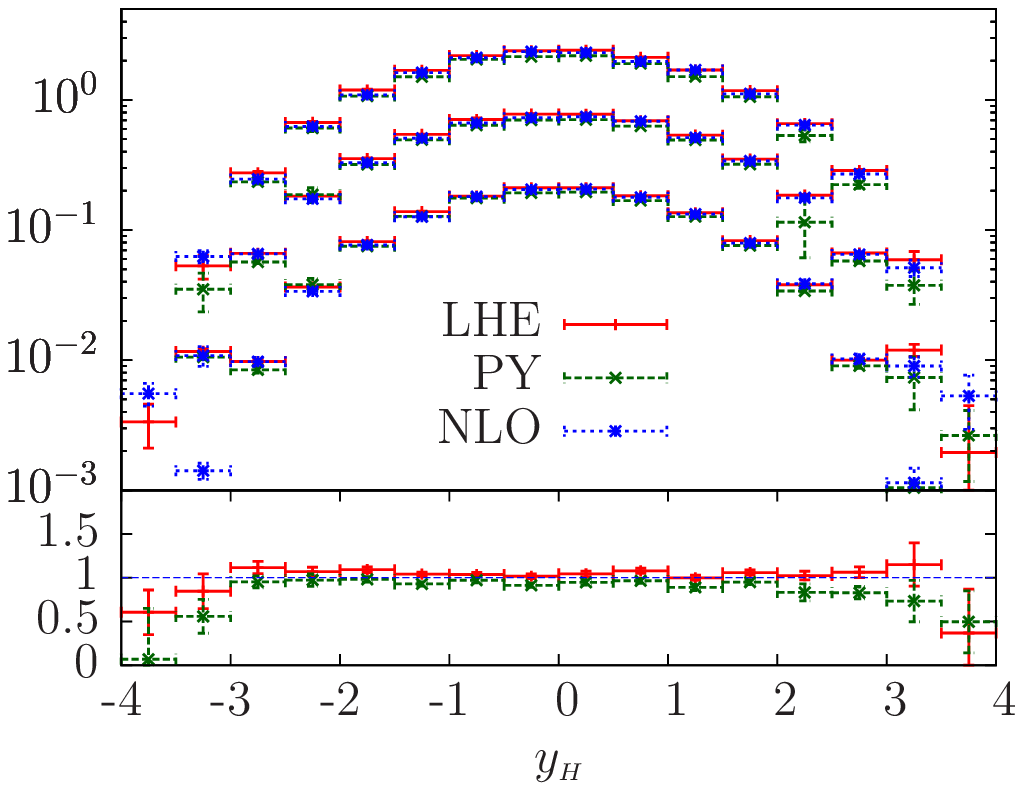} 
\caption{Rapidity distribution of the Higgs boson in the $Hj$ process. The
  three sets of curves refer to the three cuts on the hardest jet,
  i.e.~$\pt^{\sss j_1} > 20$, 50 and 100~GeV, from top to bottom respectively. }
\label{fig:HJ_H-y}
\end{center}
\end{figure}
\begin{figure}[htb]
\begin{center}
\includegraphics[height=\hfigs]{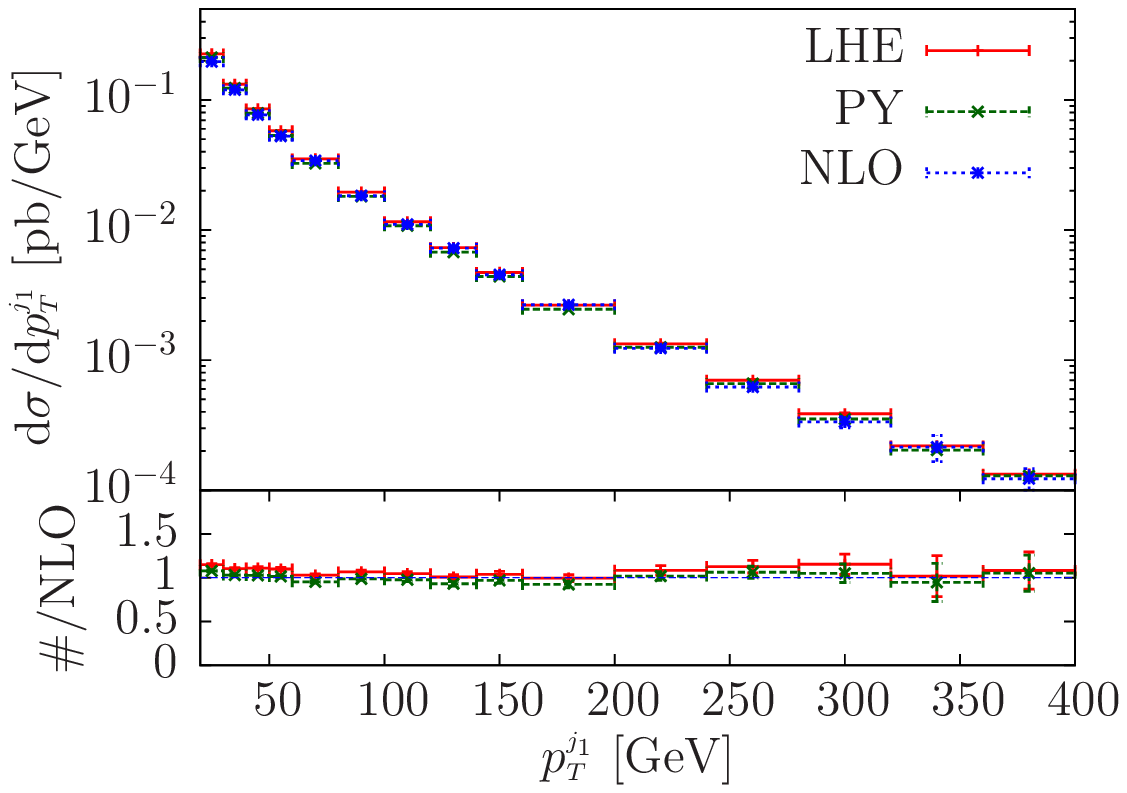} \nolinebreak
\includegraphics[height=\hfigs]{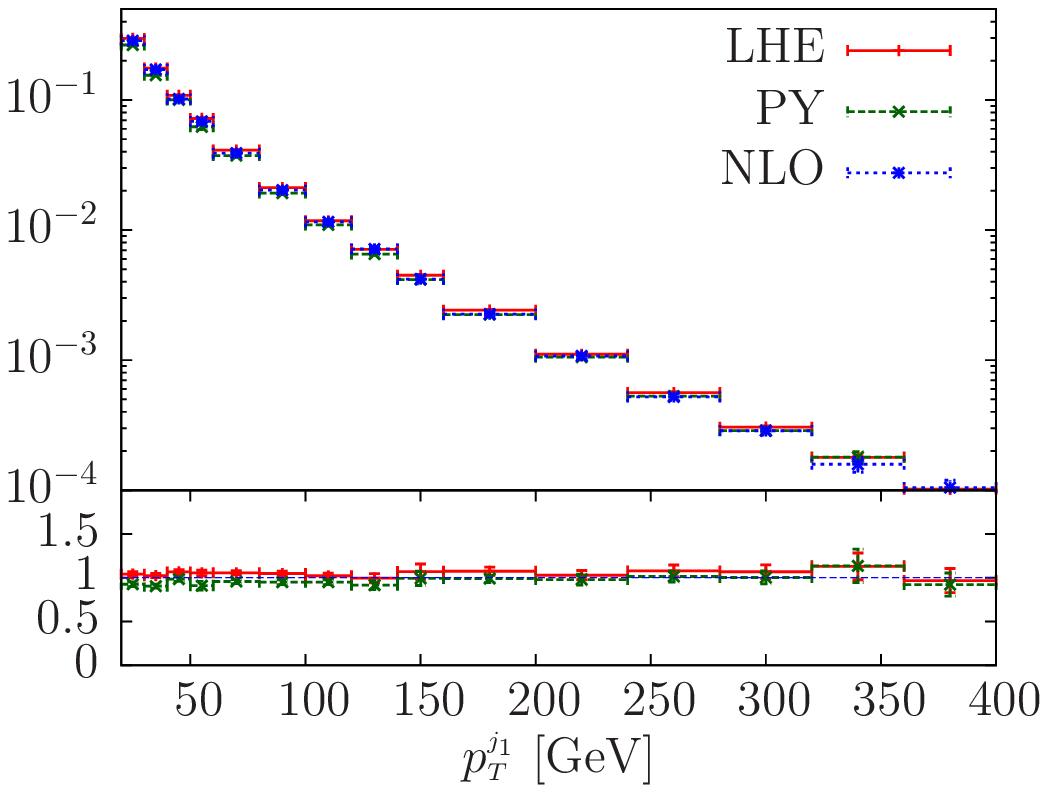} 
\caption{The hardest jet distribution in the $Hj$ process.}
\label{fig:HJ_j1-pt}
\end{center}
\end{figure}
The Higgs boson rapidity distribution is plotted in fig.~\ref{fig:HJ_H-y}
while the hardest jet $\pt$ in fig.~\ref{fig:HJ_j1-pt}. Both these quantities
display a behaviour similar to the Higgs boson transverse momentum, again
expected due to their inclusiveness.

\begin{figure}[htb]
\begin{center}
\includegraphics[height=\hfigs]{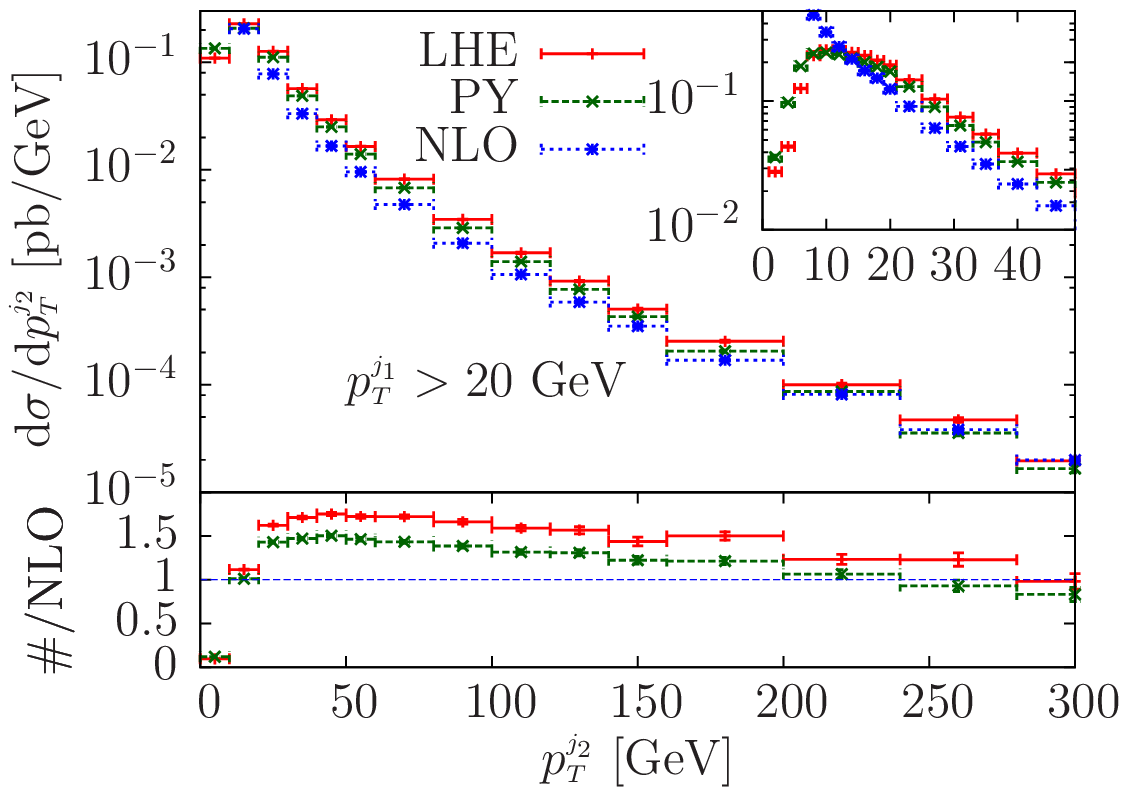}  \nolinebreak
\includegraphics[height=\hfigs]{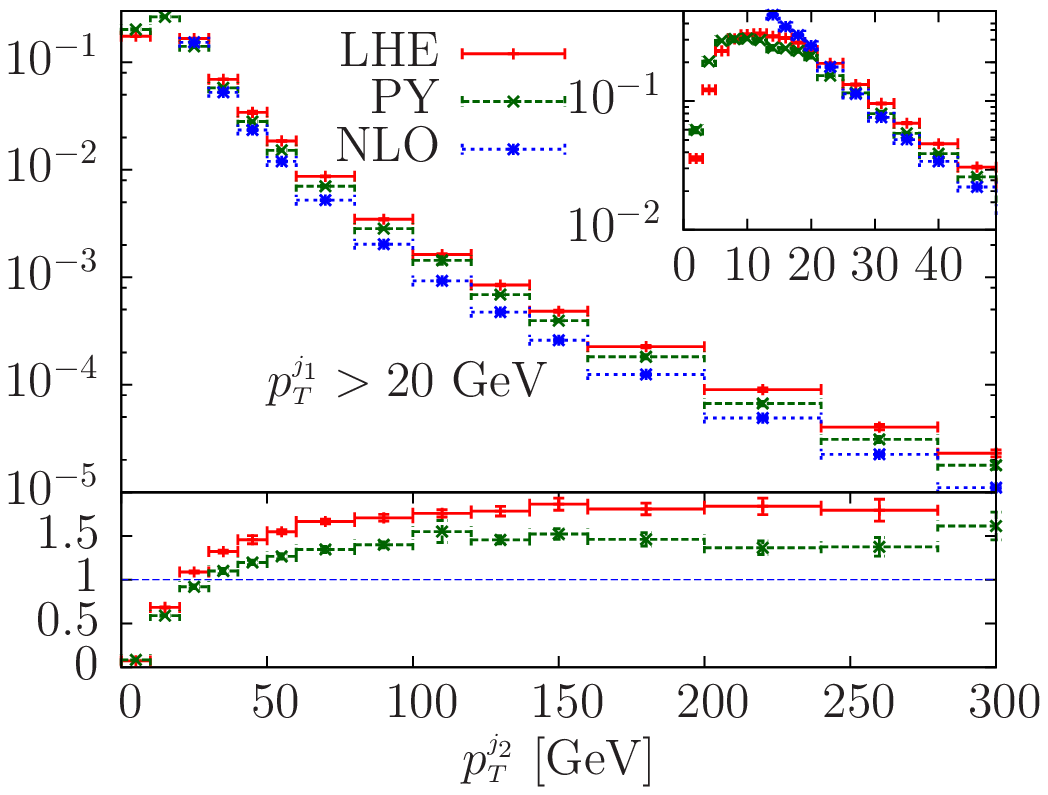} 
\caption{Transverse-momentum distribution of the second hardest jet in the
  $Hj$ process, with a 20~GeV cut on the transverse momentum of the first
  jet. Inset: the low $\pT^{\sss j_2}$ region.}
\label{fig:HJ_j2-pt}
\end{center}
\end{figure}

We now turn to more exclusive quantities, beginning with the transverse
momentum of the second jet, shown in fig.~\ref{fig:HJ_j2-pt}.
The characteristic NLO behaviour, diverging at small transverse momenta, is
clearly visible in the inset. The LHE result displays, instead, the typical
Sudakov damping at small $\pT$, compensated by an increase at larger
$\pT$. The fully showered result is smaller, since further showering degrades
the second jet transverse momentum. The two-jet multiplicity, displayed in
fig.~\ref{fig:HJ_mult}, is consistent with the value of the cross section
around $\pT=20\;$GeV in fig.~\ref{fig:HJ_h-pt}.  From the upper inset in the
figure, we see that, at this value of transverse momentum, the LHE cross
section is above the NLO one for the fixed-scale choice (left plot), while it
is near to it for the running-scale choice (right plot). For larger $\pT$
cuts, the LHE cross section remains above the NLO one in both cases, and the
ratio grows in the running-scale case.  The comparison of the fixed- and
running-scale choice in the figure also has some subtle features that should
be remarked upon. The running-scale choice yields smaller/larger scales for
smaller/larger second jet $\pT$. The LO cross section is larger for smaller
scale choices, so that the NLO corrections are smaller, and vice-versa for
larger scale choices. Thus, for small $\pT$, the NLO corrections are smaller
than for large $\pT$.  In \POWHEG{}, one first generates the underlying-Born
configuration, according to the cross section $\bar{B}$, which is the total
inclusive cross section at a fixed underlying-Born configuration. The
radiation kinematics is then generated using a shower technique. As a result
the whole distribution for the radiation is amplified by a $K$-factor equal
to $\bar{B}/B$~\cite{Alioli:2008tz, Nason:2010ap, Nason:2012pr}.  In this
case, due to the growth of the NLO corrections as a function of $\pT$, the
amplification of the radiated-jet distribution due to the $\bar{B}/B$
$K$-factor increases as a function of the transverse momentum, a trend that
is visible in the right figure. Conversely, no such effect is present for
fixed scales. However, in this last case, one should recall that, in the LHE
events, one power of $\as$ is effectively evaluated at the transverse
momentum of the jet rather than at the fixed scale, thus yielding a decrease
in the cross section which is also visible in the left plot.
The larger value of the LHE cross section with respect to the NLO one is
related to the large $K$-factor, i.e.~is due to the fact that the hardest
radiation is amplified by a factor $\bar{B}/B$.

\begin{figure}[htb]
\begin{center}
\includegraphics[height=\hfigs]{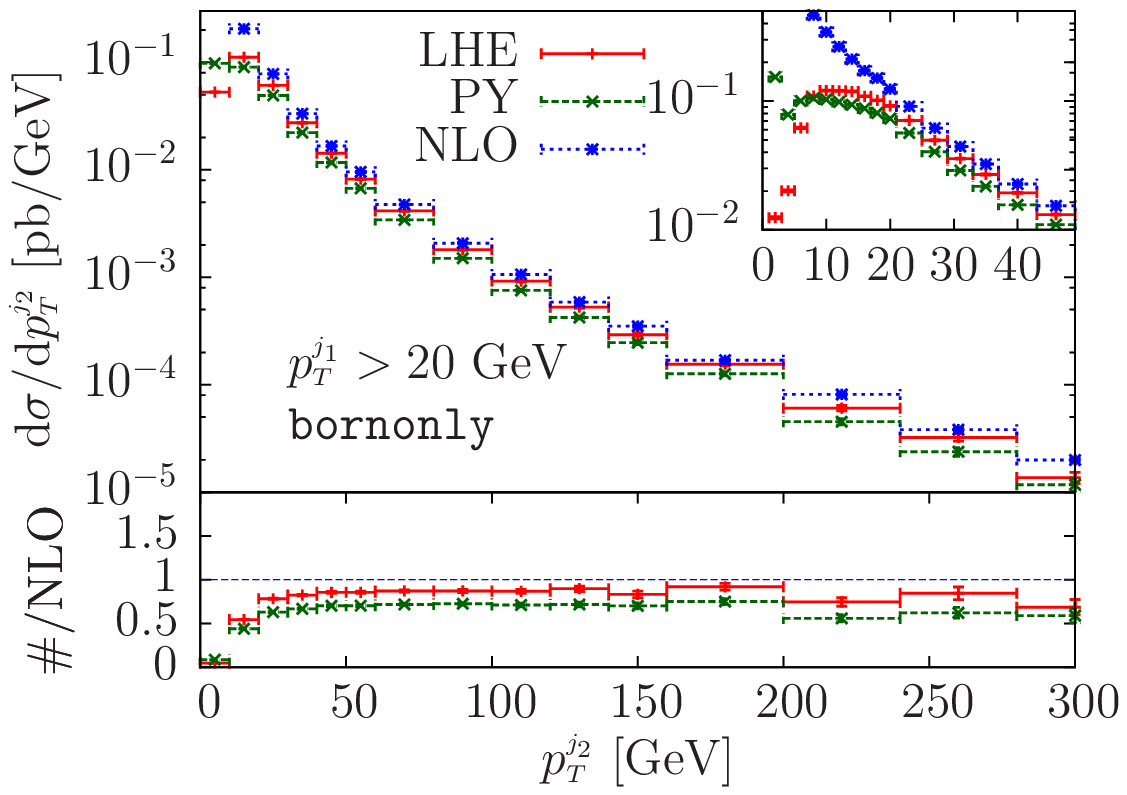} \nolinebreak
\includegraphics[height=\hfigs]{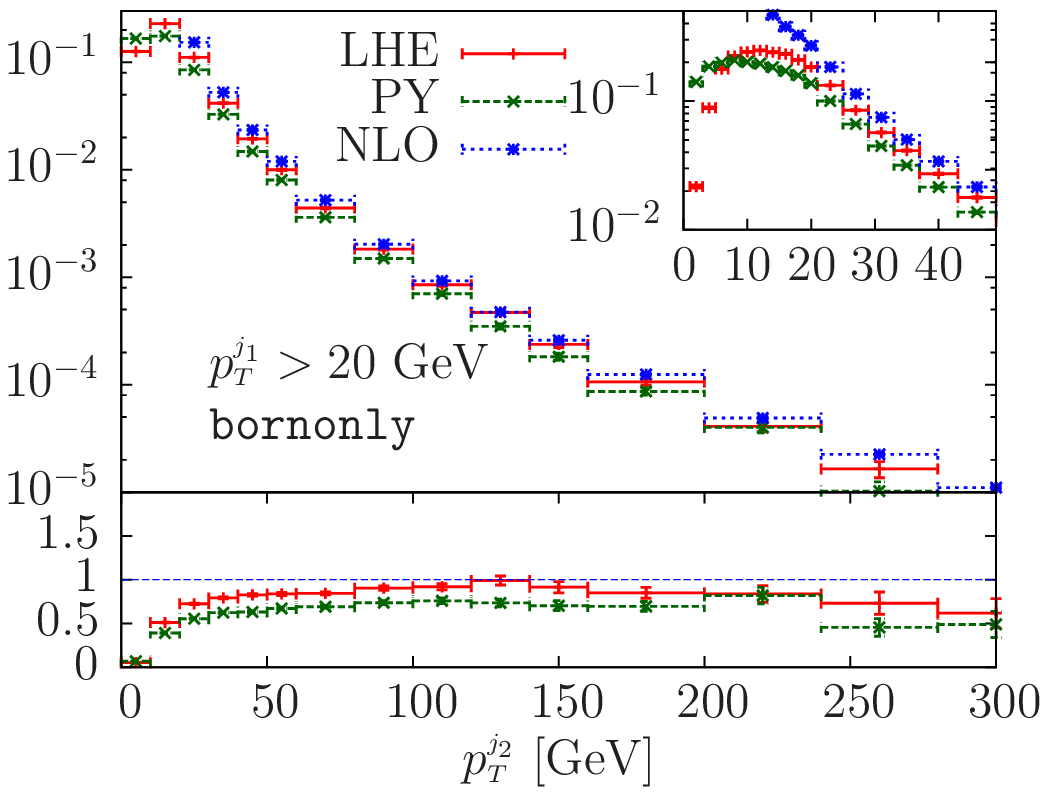} 
\caption{Transverse-momentum distribution of the second hardest jet using the
  $\bar{B}\to B$ option.}
\label{fig:HJ_j2-pt_bornonly}
\end{center}
\end{figure}
Figure~\ref{fig:HJ_j2-pt_bornonly} is similar to fig.~\ref{fig:HJ_j2-pt}
except that the LHE result is obtained setting the {\tt bornonly} flag to
true in the \POWHEGBOX{} input file. When this flag is true, the \POWHEGBOX{}
generator uses the Born cross section, rather than the NLO $\bar{B}$
function, to generate the underlying-Born configuration. We can see that, in
this case, the LHE result is much closer to the NLO result, except for small
transverse momenta, where the Sudakov damping becomes manifest. This results
confirms that the enhancement of the transverse-momentum distribution of the
second hardest jet is indeed due to the $\bar{B}/B$ $K$-factor.

\begin{figure}[htb]
\begin{center}
\includegraphics[height=\hfigs]{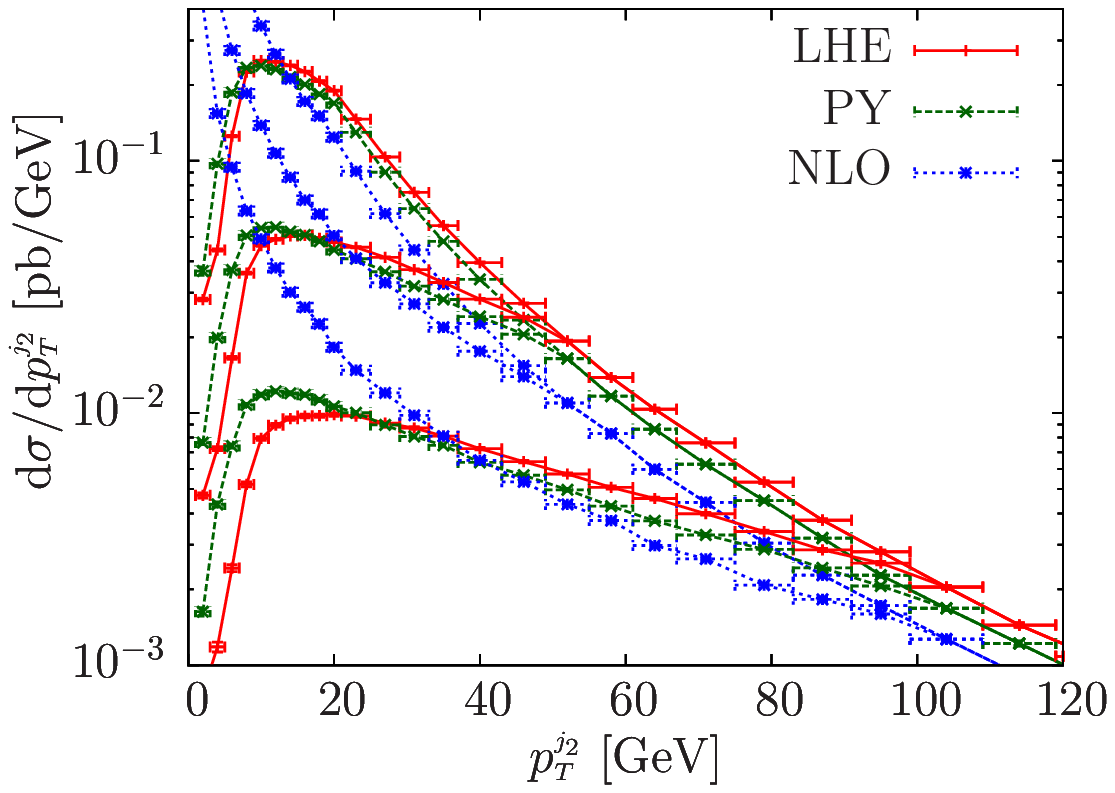}\nolinebreak
\includegraphics[height=\hfigs]{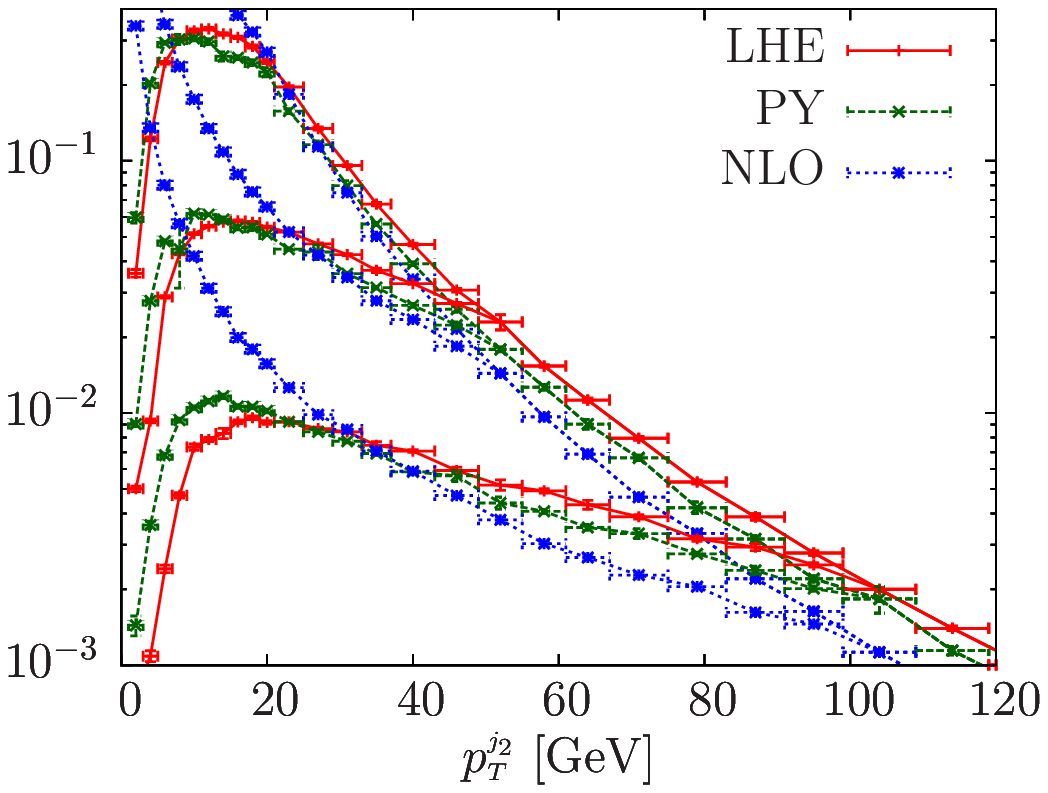}
\caption{Transverse-momentum distribution of the second hardest jet for
  different cuts on the first jet transverse momentum.}
\label{fig:HJ_j2-pt_cuts}
\end{center}
\end{figure}
In fig.~\ref{fig:HJ_j2-pt_cuts} we show the second hardest jet transverse
momentum for different cuts on the first jet $\pT$. From the figure it is
clear that, when the second jet has transverse momentum above the first jet
$\pT$ cut, the leading jet is forced to have larger transverse momentum, and
the cross section falls more rapidly. The comparison among the LHE, PY and
NLO curves shows the typical pattern, with the NLO diverging at small
transverse momenta, the LHE being instead suppressed in that region, but
raising above the NLO result because of the $\bar{B}/B$ $K$-factor.

\begin{figure}[htb]
\begin{center}
\includegraphics[height=\hfig]{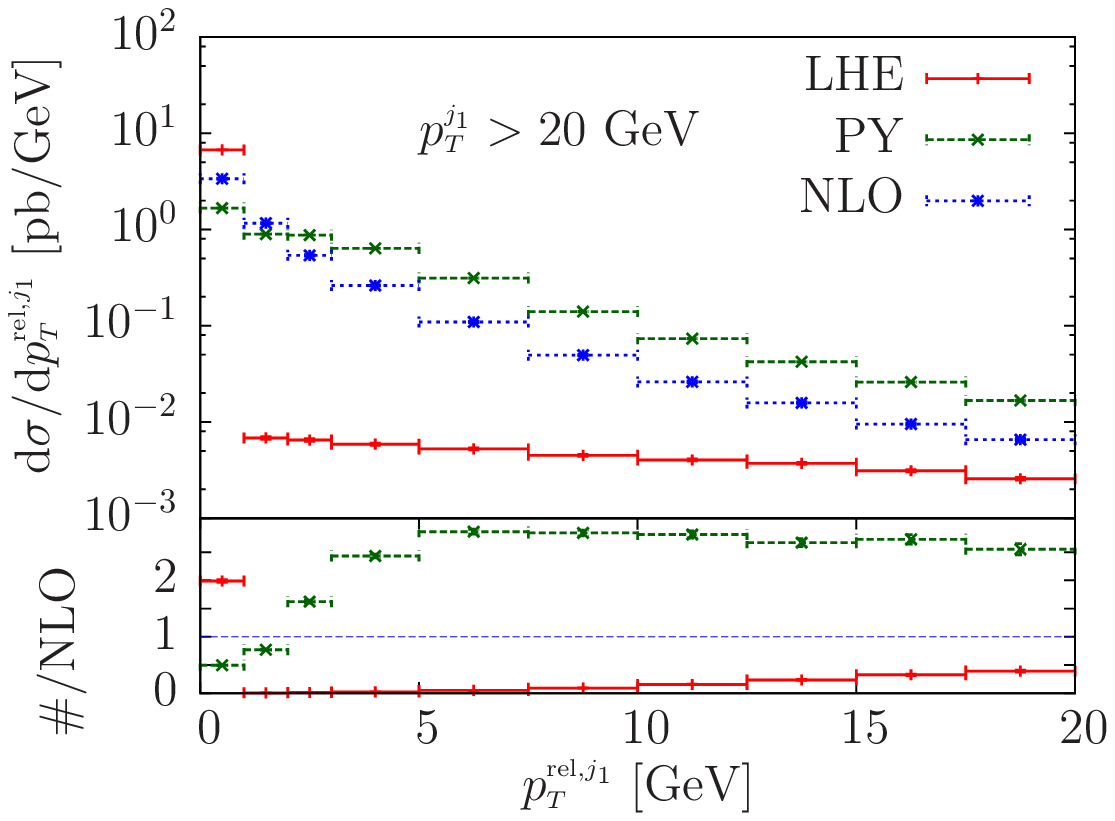}\nolinebreak
\includegraphics[height=\hfig]{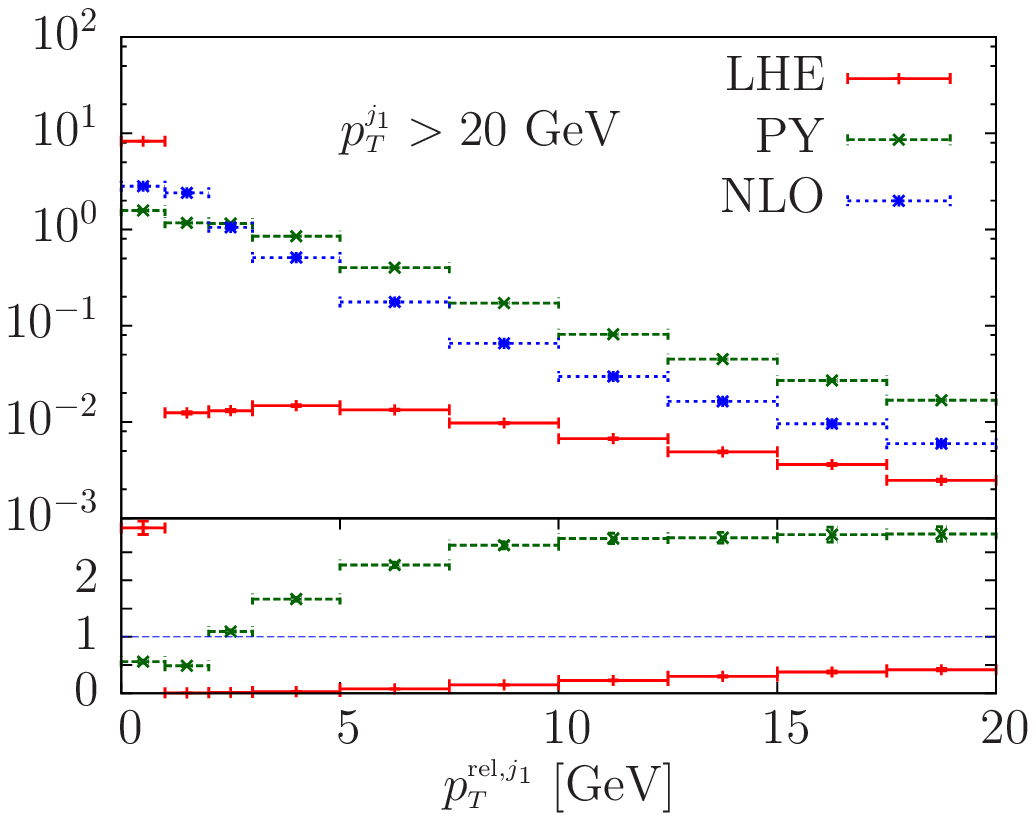}
\caption{The relative transverse momenta of the partons within the hardest
  jet $\ptrelone$ as defined in eq.~(\ref{eq:ptrel1}).}
\label{fig:HJ_ptrel1}
\end{center}
\end{figure}
In fig.~\ref{fig:HJ_ptrel1} we plot the relative transverse momenta of the
partons within the hardest jet $\ptrelone$.  This quantity is defined as
follows: we perform a longitudinal boost to a frame where the rapidity of the
first jet $j_1$ is zero. In this frame we compute
\begin{equation}
\label{eq:ptrel1}
p_{\sss T}^{ \sss {\rm rel},j_1}=\sum_{i\in j_1}
\frac{|\vec{p}_i\times \vec{k}_j|}{|\vec{k}_j|}\,,
\end{equation}
where $k_j$ is the momentum of the first jet, and $p_i$ the momenta of the
partons clustered within the first jet.  This quantity, when computed at the
LHE level, displays a marked difference from the corresponding NLO result. As
discussed in ref.~\cite{Alioli:2010xa}, this can be easily understood if we
remember that the LHE result is suppressed by a Sudakov form factor, which
requires that no harder radiation has been emitted from either the initial-
or the final-state partons.  The large bin at $\ptrelone=0$ in the LHE result
is due to events in which there is only one parton in the jet, most likely
initial-state radiation events. On the other hand, the hardest radiation
provided by \PYTHIA{} should restore a shape closer to the NLO result,
although amplified by the $\bar{B}/B$ factor.  Further amplification of the
showered result is due to the fact that the shower uses a running coupling
evaluated at the scale of the radiation, while the NLO result uses higher
scales, and to the presence of multiple emissions in the shower.  It is clear
that this distribution, being determined mostly by the shower program, is
quite sensitive to the shower model and tuning, and to the interface between
the shower and \POWHEGBOX{}.

\subsection{Results for $Hjj$ production}
We have generated a sample with 2.5M events both at fixed and running scales.
The event generation time is approximately 25 minutes for 1000 events on a
typical CPU.

We have run the $Hjj$ program for two scale choices, $\muf=\mur=\mH$
and $\muf=\mur=\HThat$. All the following plots will come in pairs,
the left one referring to the first scale choice, and the right one referring to
the second scale choice.

\begin{figure}[htb]
\begin{center}
\includegraphics[height=\hfig]{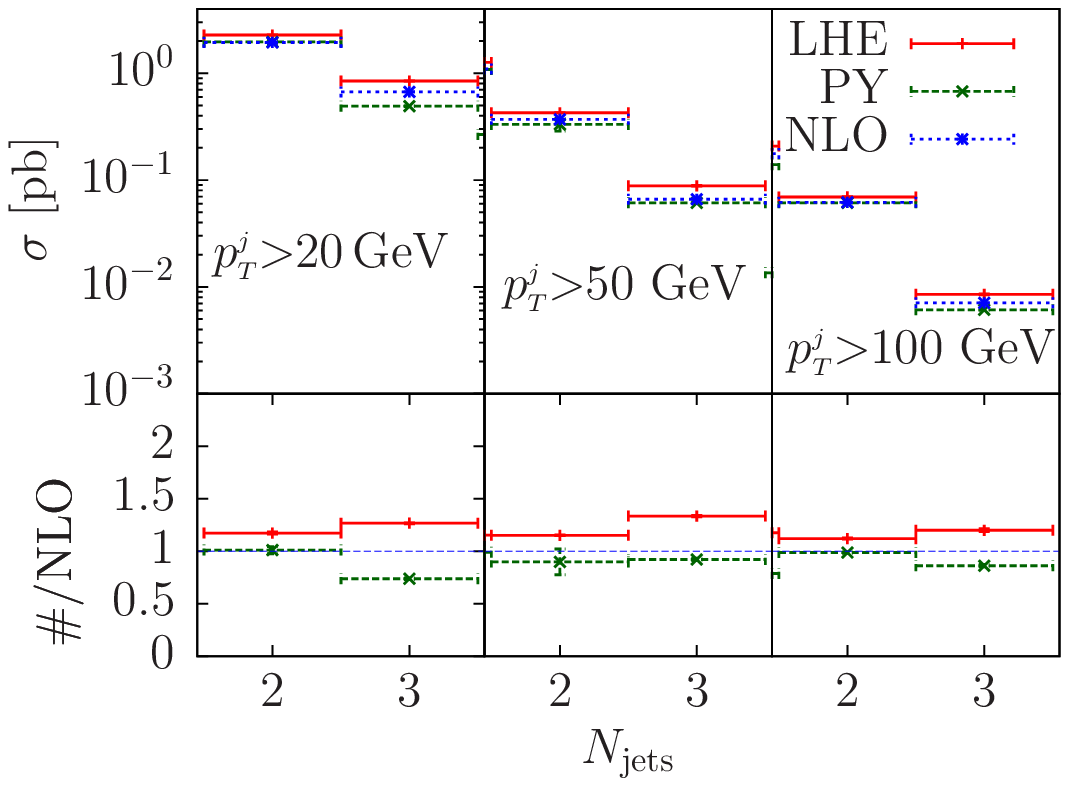} \nolinebreak
\includegraphics[height=\hfig]{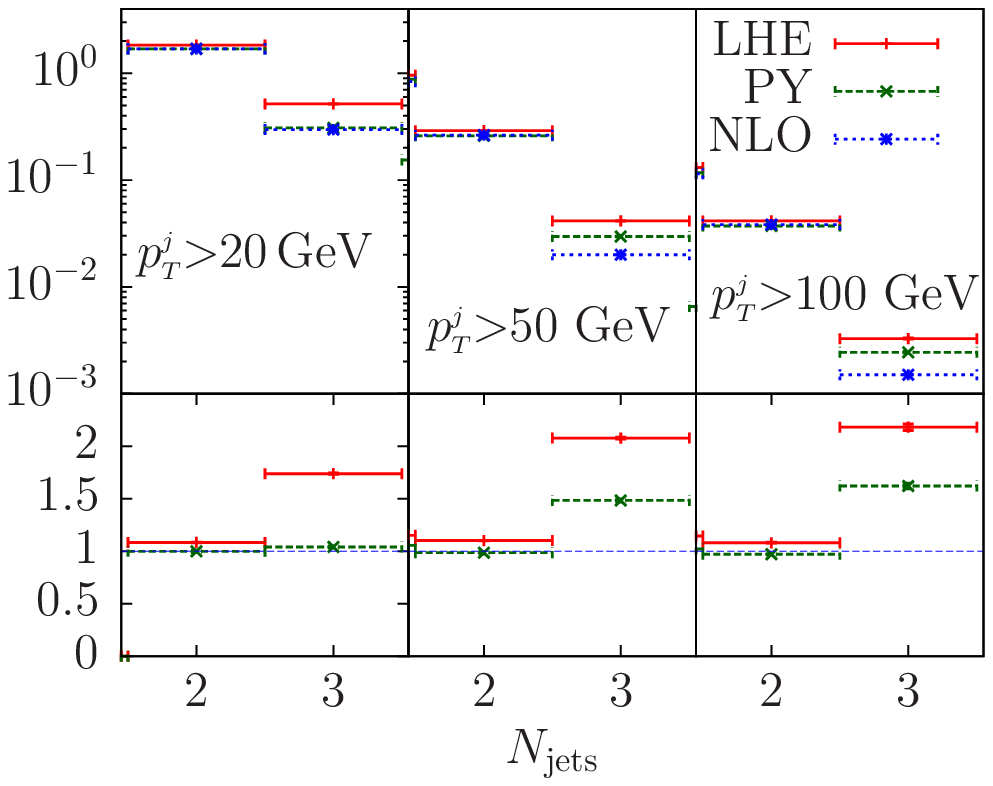} 
\caption{Two- and three-jet multiplicities in $Hjj$ production. In the left plot
  the scale is chosen equal to the mass of the Higgs. In the right plot the
  scale is taken equal to $\HThat$.}
\label{fig:njjmult}
\end{center}
\end{figure}
We begin by showing in fig.~\ref{fig:njjmult} the two- and three-jet
multiplicities in the $Hjj$ process.  We find considerable agreement of the
LHE, PY and NLO output for both the two- and three-jet multiplicities for the
left plot, which corresponds to the choice of scale $\muf=\mur=\mH$. The
right plot corresponds to the $\HThat$ choice of scale. In this plot the
two-jet multiplicity agrees for all cuts, while the three-jet multiplicity
displays marked differences between the LHE results and the NLO ones. For the
$100$ GeV transverse-momentum cut, the fully showered result and the NLO also
differ. The reason for these differences is the following: with the $\mH$
scale choice the $K$-factor is near one, and thus the $\bar{B}/B$
amplification that usually enhances the spectrum of the radiated jet (that in
this case is the third jet) is not effective. The $\HThat$ scale is
considerably larger than $\mH$, especially with a large $\pT$ cut, which
implies a reduction of the Born cross section and an increase of the
$K$-factor.

\begin{figure}[htb]
\begin{center}
\includegraphics[height=\hfigs]{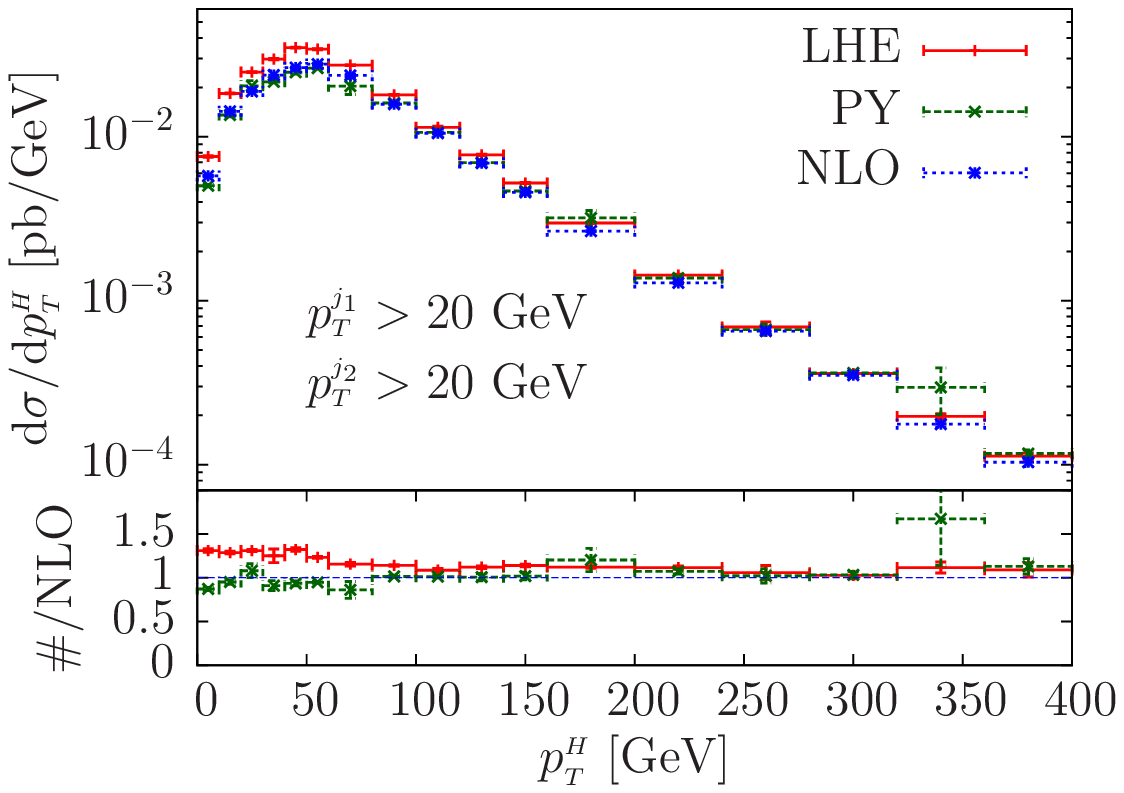} \nolinebreak
\includegraphics[height=\hfigs]{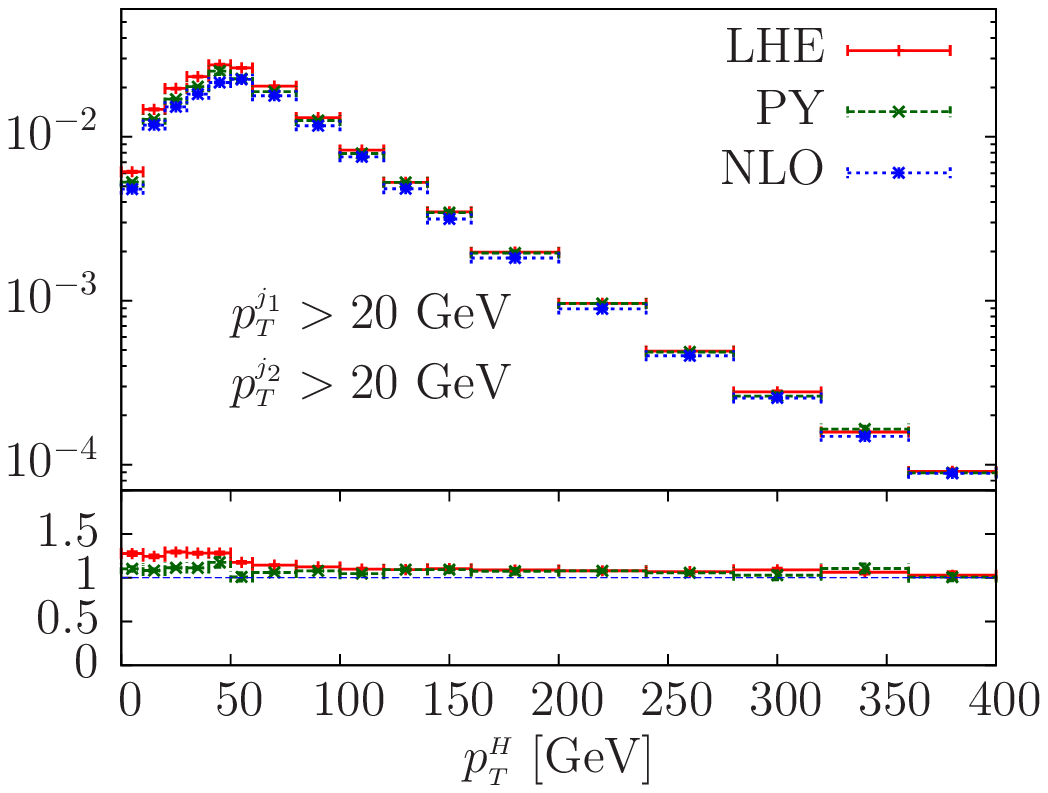} 
\caption{Transverse-momentum distribution of the Higgs boson in a $Hjj$
  sample with a 20~GeV $\pT$ cut on the two hardest jets. The left and right
  plots use respectively the $\mH$ and $\HThat$ scales.}
\label{fig:HJJ_H-pt}
\end{center}
\end{figure}
In fig.~\ref{fig:HJJ_H-pt} we show the transverse-momentum distribution of
the Higgs boson in a $Hjj$ sample with a 20~GeV $\pT$ cut on the two hardest
jets.  Good agreement is found between the LHE, PY and NLO curves, except for
small Higgs transverse momenta, where differences of the order of 30\% are
found. This is not surprising, in view of the required presence of two
relatively soft jets. Radiation off these jets is Sudakov suppressed in the
LHE result with respect to the NLO one.  Since this radiation would deplete
the jets, the LHE result suffers less depletion, and is thus larger. On the
other hand, the shower degrades the energy of the jets lowering the cross
section, an effect visible in the PY result. For large Higgs boson $\pT$, at
least one of the jets is forced to be hard, and thus these effects lose
importance.

\begin{figure}[htb]
\begin{center}
\includegraphics[height=\hfigs]{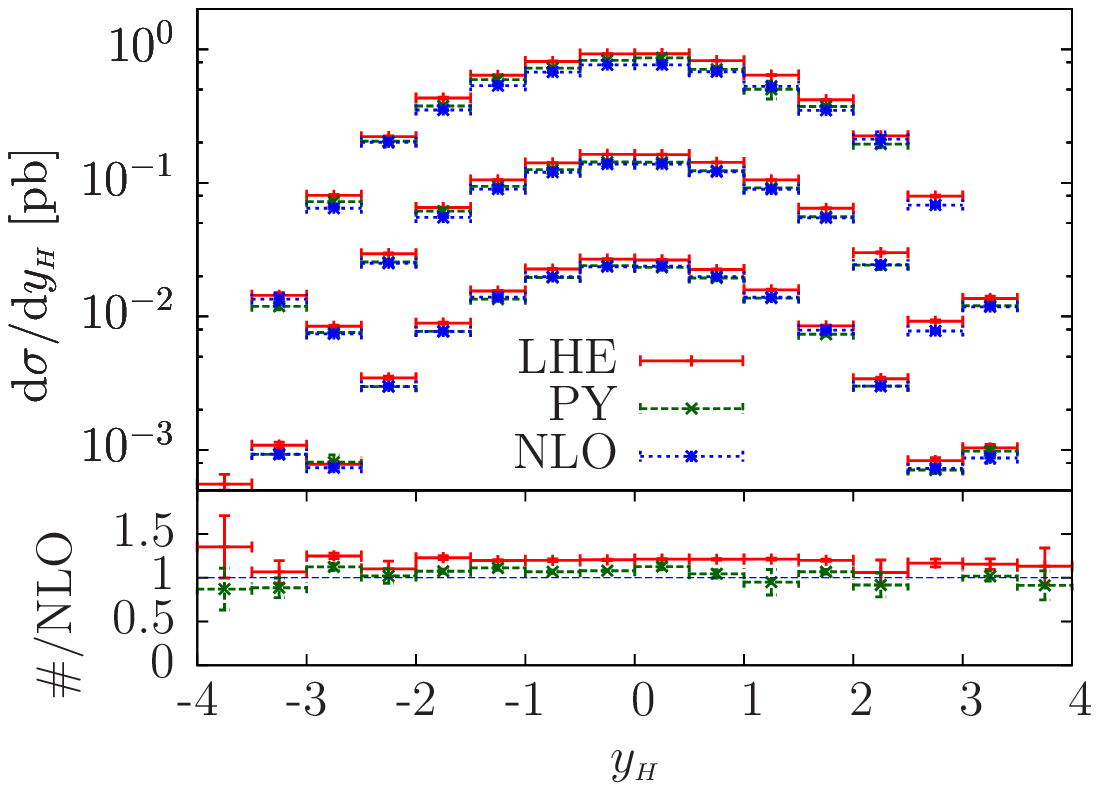}  \nolinebreak
\includegraphics[height=\hfigs]{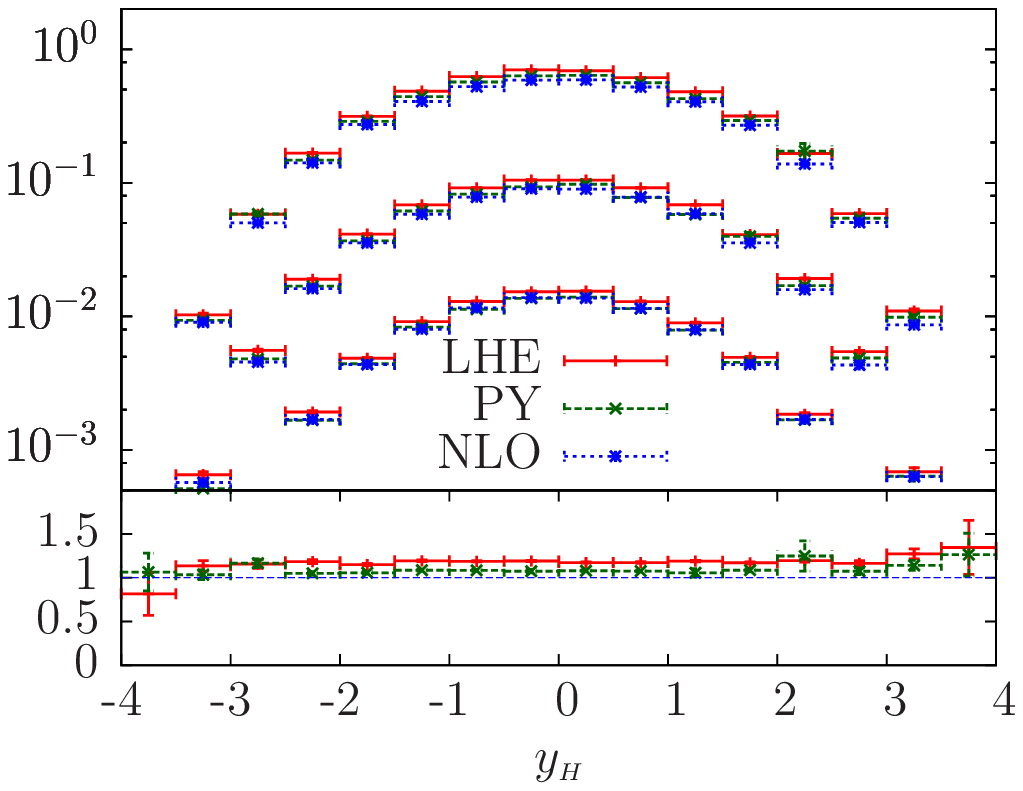} 
\caption{Higgs boson rapidity distribution in $Hjj$ for $\pT$ cuts of $20$,
  $50$~ and $100$~GeV on the hardest jets. The left and right plots use
  respectively the $\mH$ and $\HThat$ scales.  In the lower pane, the ratio
  of the LHE and PY results with respect to the NLO one, for the $20$~GeV
  cut.}
\label{fig:HJJ_H-y}
\end{center}
\end{figure}
In fig.~\ref{fig:HJJ_H-y} we show the Higgs boson rapidity distribution in
$Hjj$ for $\pT$ cuts of $20$, $50$ and $100$~GeV on the two hardest
jets. Again, since this is an inclusive distribution, it displays good
agreement between the LHE, PY and NLO results. The ratio is only displayed
for the $20\;$GeV cut. The three predictions become even more consistent for
larger $\pT$ cuts, as one would expect.

\begin{figure}[htb]
\begin{center}
\includegraphics[height=\hfigs]{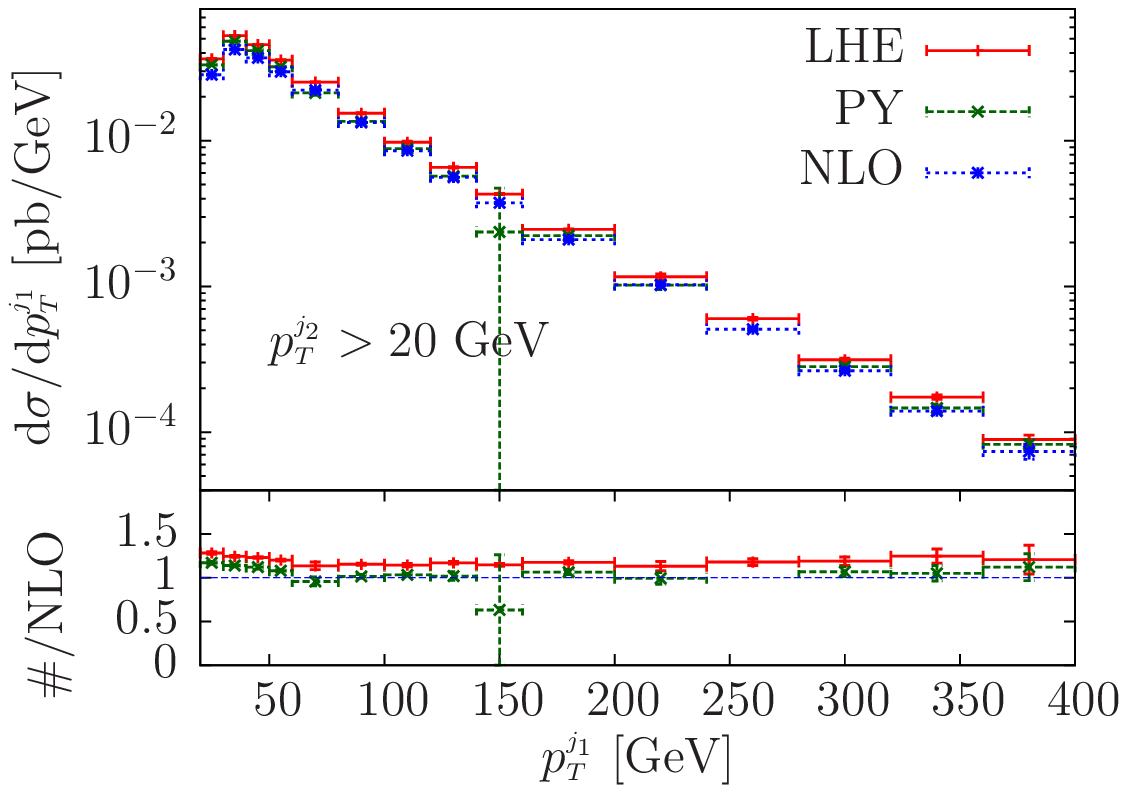}  \nolinebreak
\includegraphics[height=\hfigs]{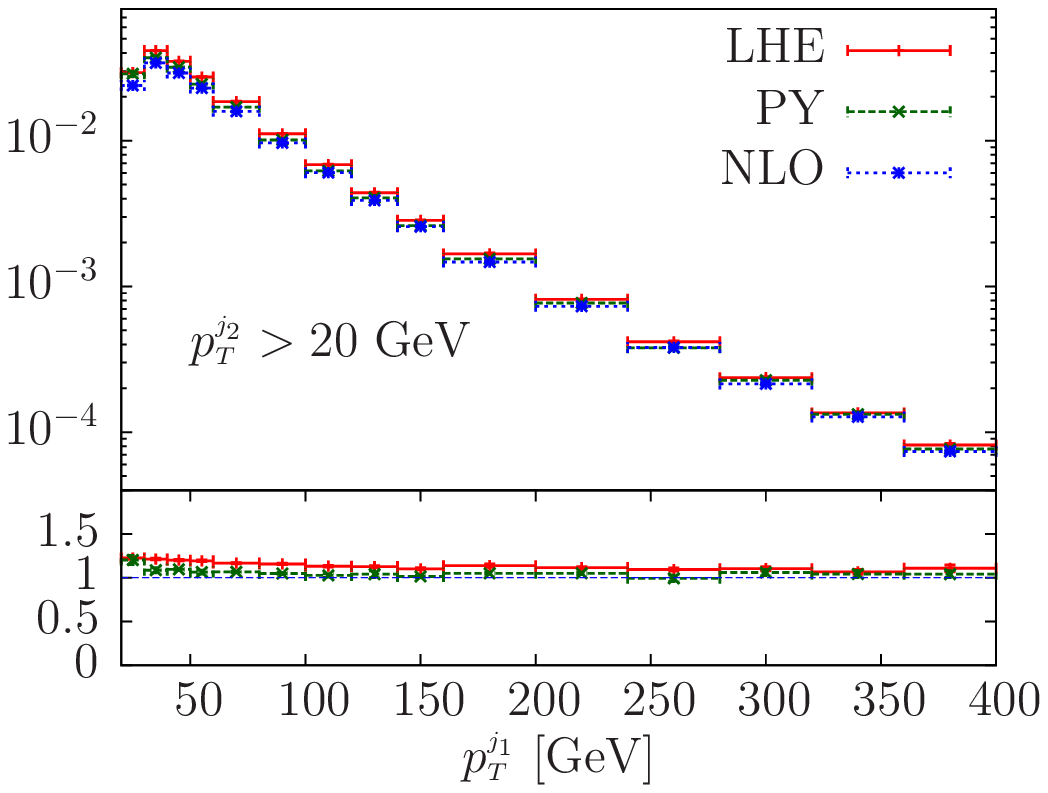} 
\caption{Transverse-momentum distribution of the hardest jet in the $Hjj$
  process. The left and right plots use respectively the $\mH$ and $\HThat$
  scales.}
\label{fig:HJJ_j1-pt}
\end{center}
\end{figure}

\begin{figure}[htb]
\begin{center}
\includegraphics[height=\hfigs]{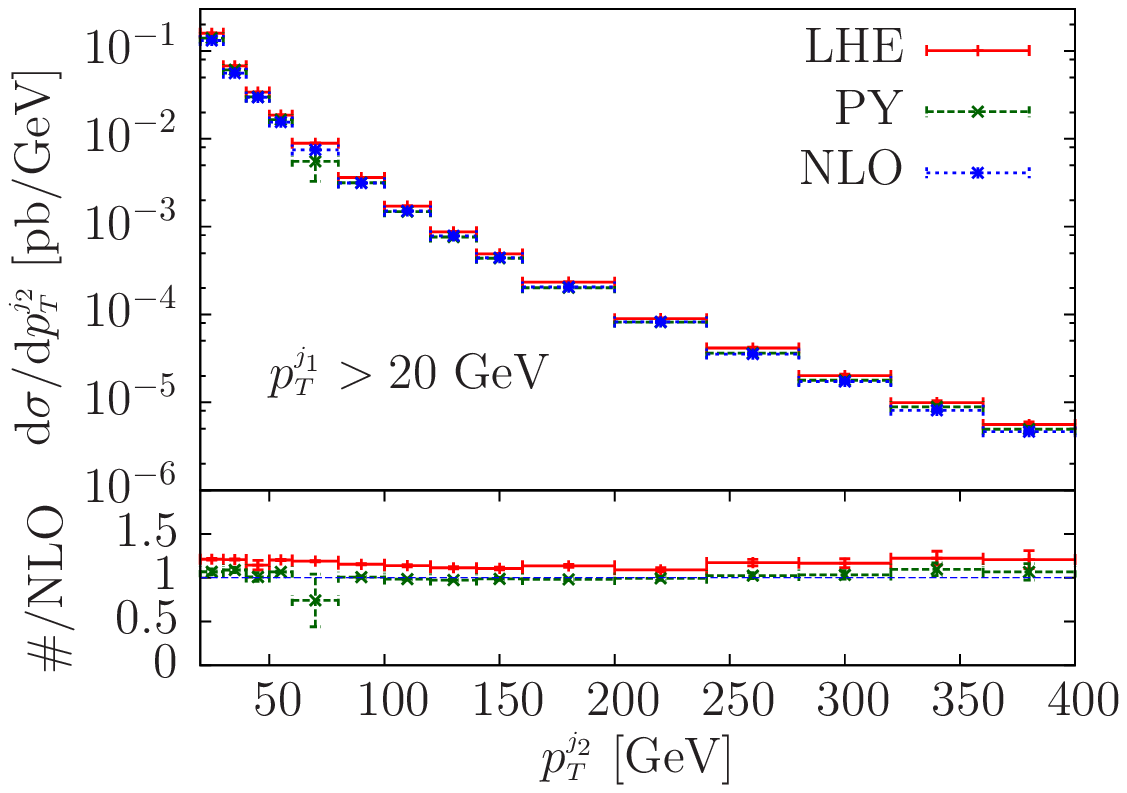}   \nolinebreak
\includegraphics[height=\hfigs]{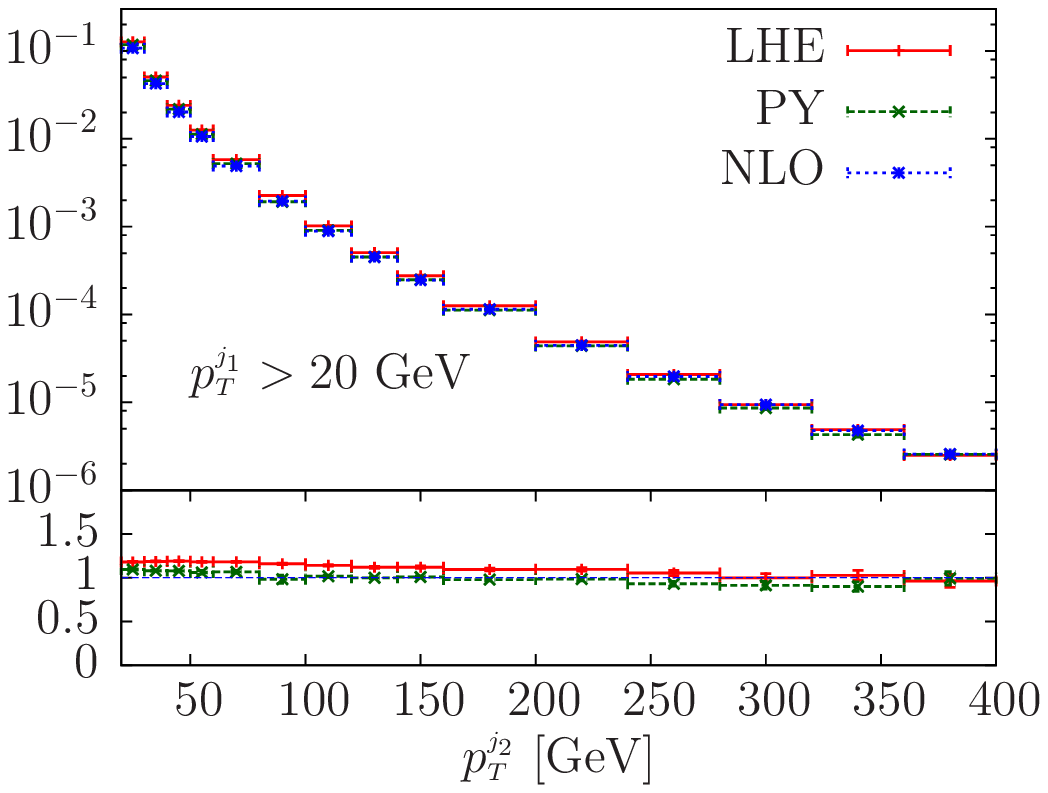} 
\caption{Transverse-momentum distribution of the second hardest jet in the
  $Hjj$ process. The left and right plots use respectively the $\mH$ and
  $\HThat$ scales.}
\label{fig:HJJ_j2-pt}
\end{center}
\end{figure}
In figs.~\ref{fig:HJJ_j1-pt} and~\ref{fig:HJJ_j2-pt} we display the
transverse-momentum distribution for the first and second jet. The first
distribution bears some similarity to the Higgs transverse momentum, for
inclusiveness reasons. Observe that the small momentum disagreement between
the LHE and the NLO result observed for the Higgs transverse momentum case is
less evident in the hardest jet plot, and even less so in the second jet
spectrum. This is explained by the fact that the same point in the abscissa
corresponds to an increasing hardness of the event in the Higgs boson, first
jet and second jet $\pT$ distributions.

\begin{figure}[htb]
\begin{center}
\includegraphics[height=\hfigs]{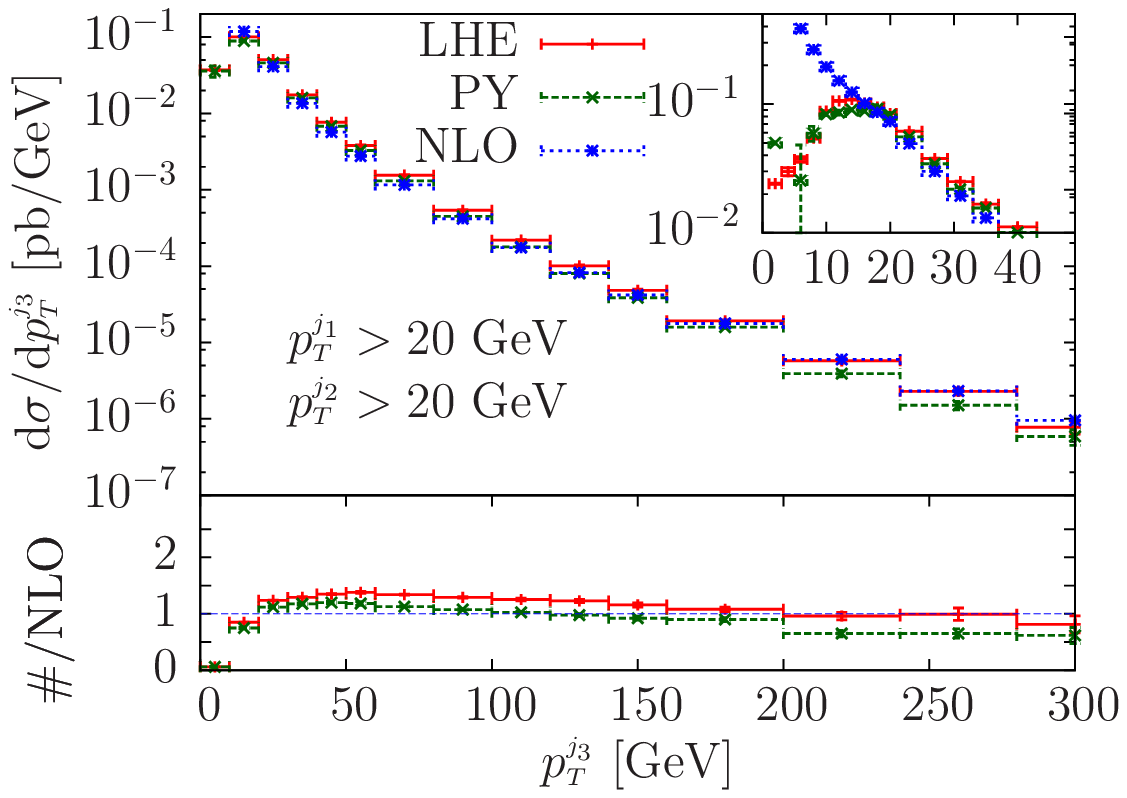}  \nolinebreak
\includegraphics[height=\hfigs]{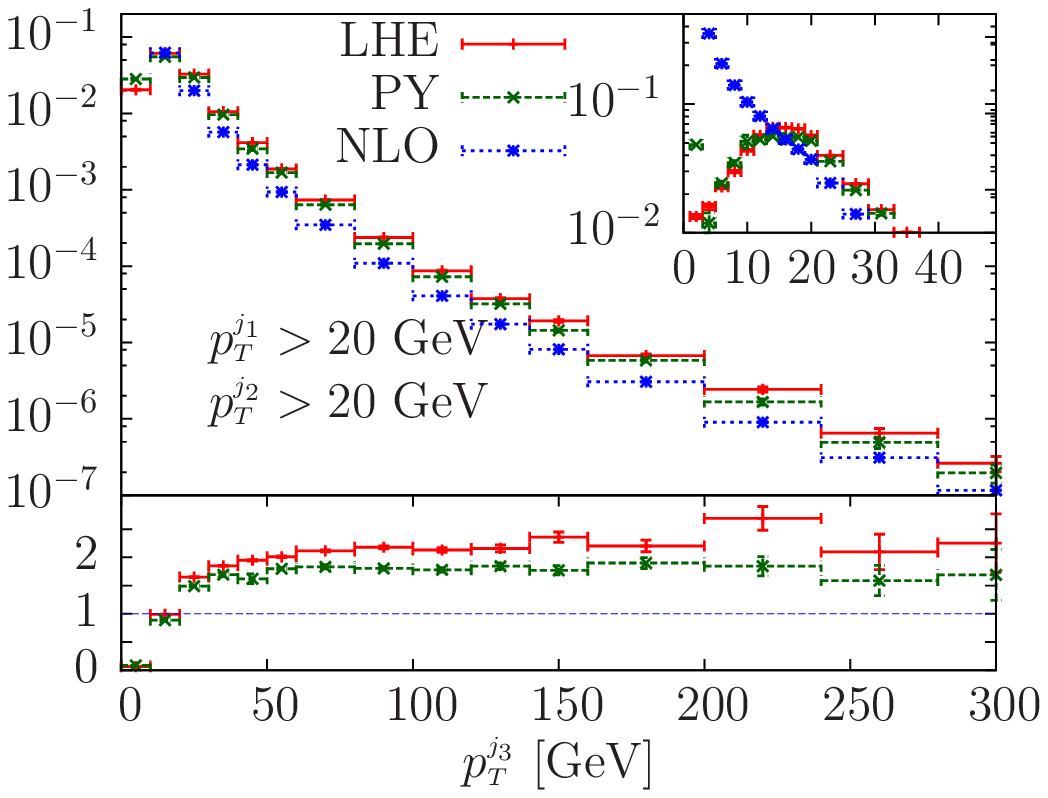} 
\caption{Transverse-momentum distribution of the third hardest jet in the
  $Hjj$ process. The left and right plots use respectively the $\mH$ and
  $\HThat$ scales.}
\label{fig:HJJ_j3-pt}
\end{center}
\end{figure}
Turning to less inclusive quantities, we show in fig.~\ref{fig:HJJ_j3-pt} the
transverse-momentum distribution of the third hardest jet.  We recognize here
the typical behaviour of the LHE results, with the damping of the
small-momentum growth found in the NLO result, and with the typical increase
due to the $\bar{B}/B$ factor at higher momenta.  As already anticipated in
the discussion on the jet multiplicity, we notice that the increase is
modest for the $\mH$ choice of scale, but is very large for the
$\HThat$ scale.

\begin{figure}[htb]
\begin{center}
\includegraphics[height=\hfigs]{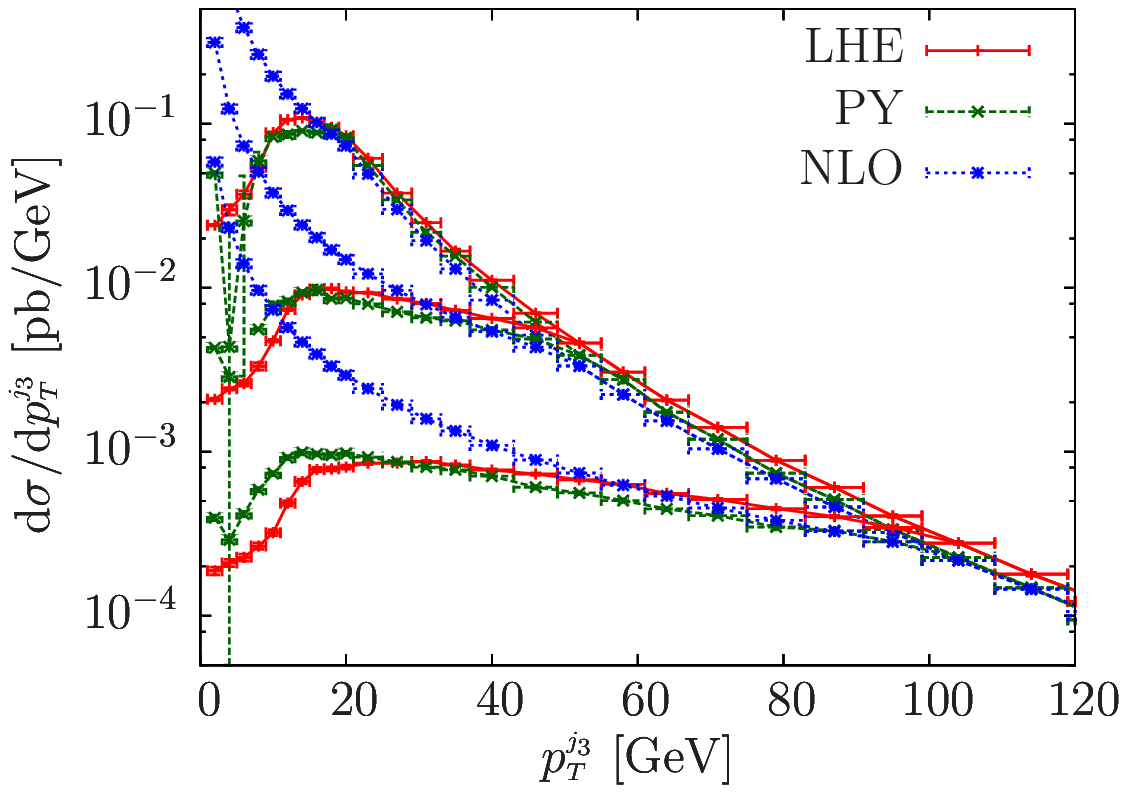}\nolinebreak  
\includegraphics[height=\hfigs]{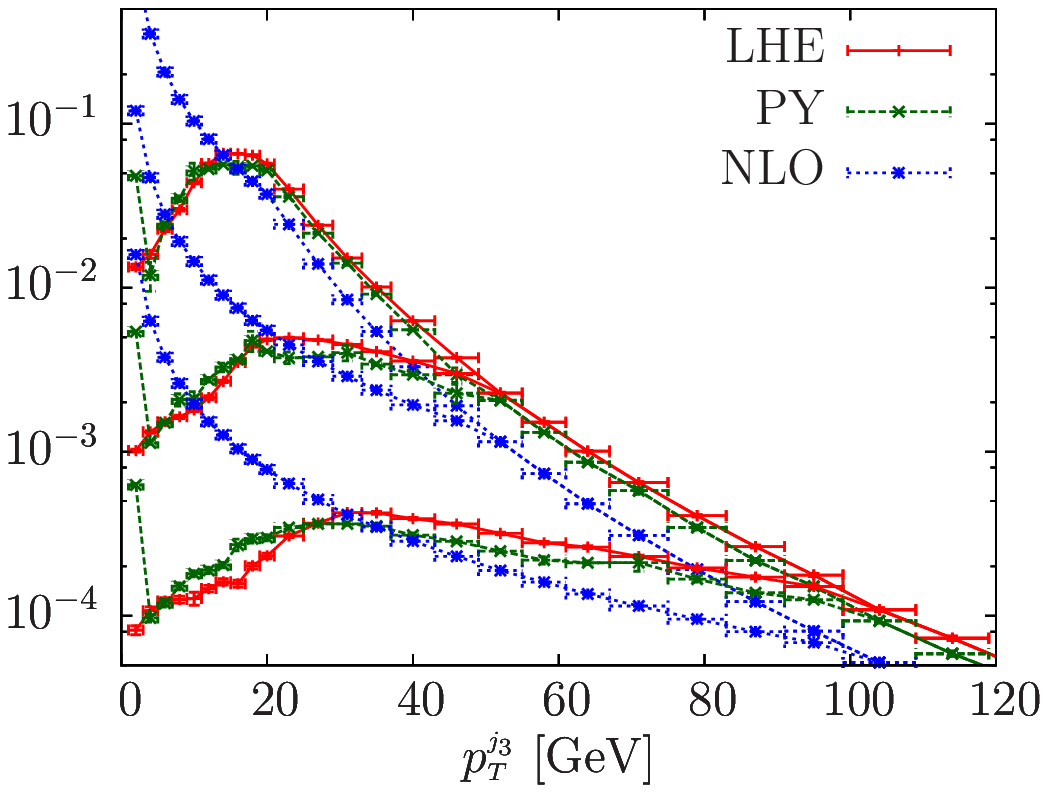} 
\caption{Transverse-momentum distribution of the third hardest jet in the
  $Hjj$ process for the transverse momentum cut on the two hardest jets equal
  to $20$, $50$ and $100$~GeV.}
\label{fig:HJJ_j3-pt_cuts}
\end{center}
\end{figure}
In fig.~\ref{fig:HJJ_j3-pt_cuts} we display the third-jet transverse momentum
using three different cuts on the $\pT$ of the first and second jet. The
pattern is similar to what was already observed in the $Hj$ case. In the left
plot, where the scale is chosen equal to $\mH$, we see a better concordance
of the LHE and NLO results as the transverse momentum increases.

\begin{figure}[htb]
\begin{center}
\includegraphics[height=\hfigs]{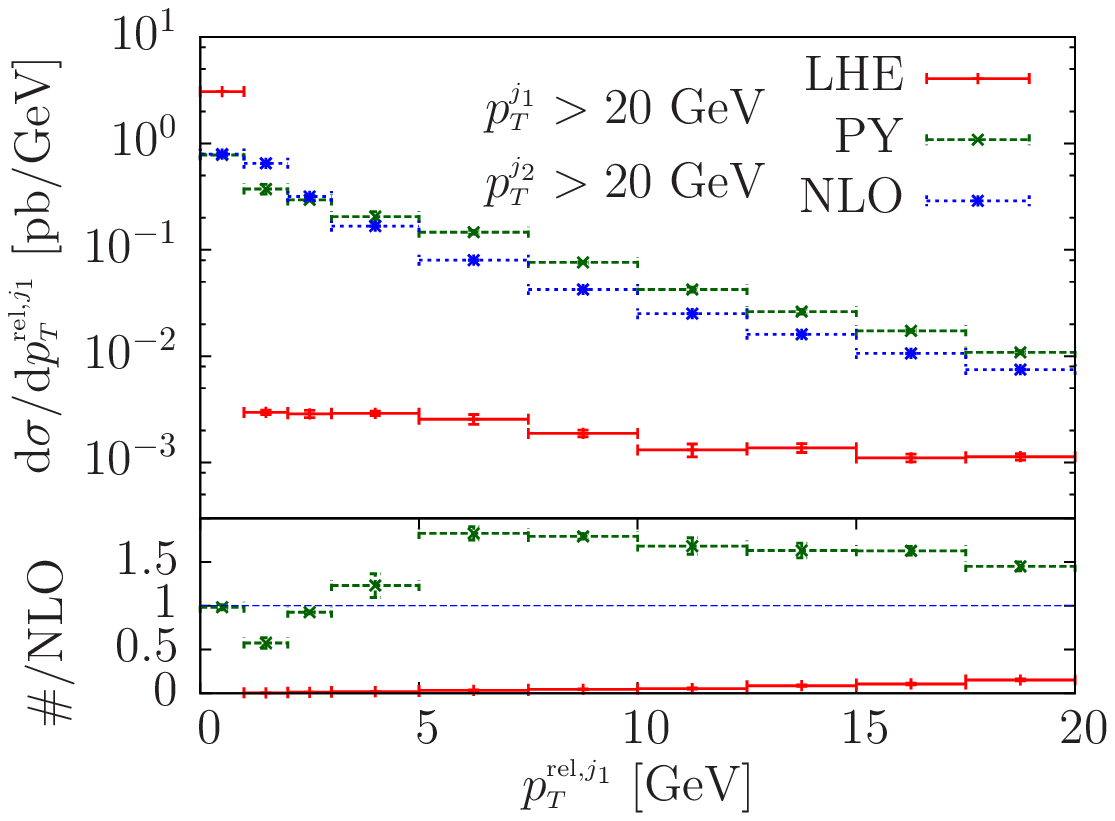} \nolinebreak
\includegraphics[height=\hfigs]{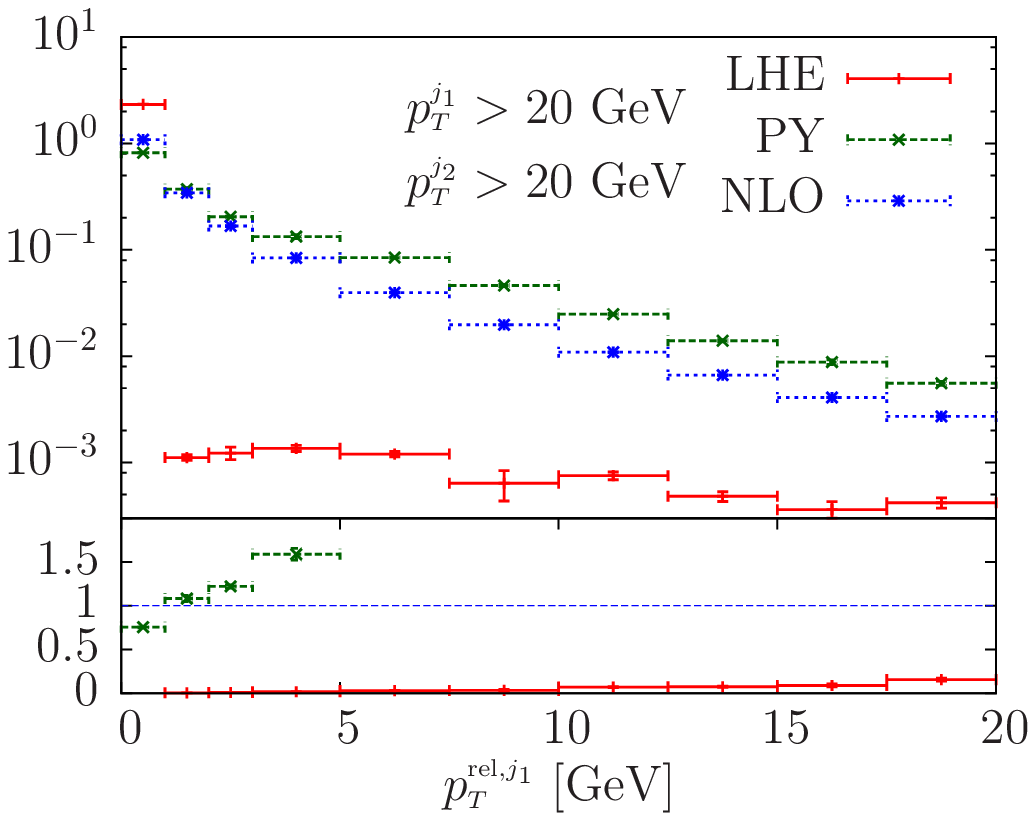} 
\caption{$\ptrelone$ distribution of the hardest jet in the $Hjj$
  process. The left and right plots use respectively the $\mH$ and $\HThat$
  scales.}
\label{fig:HJJ_ptrel1}
\end{center}
\end{figure}

\begin{figure}[htb]
\begin{center}
\includegraphics[height=\hfigs]{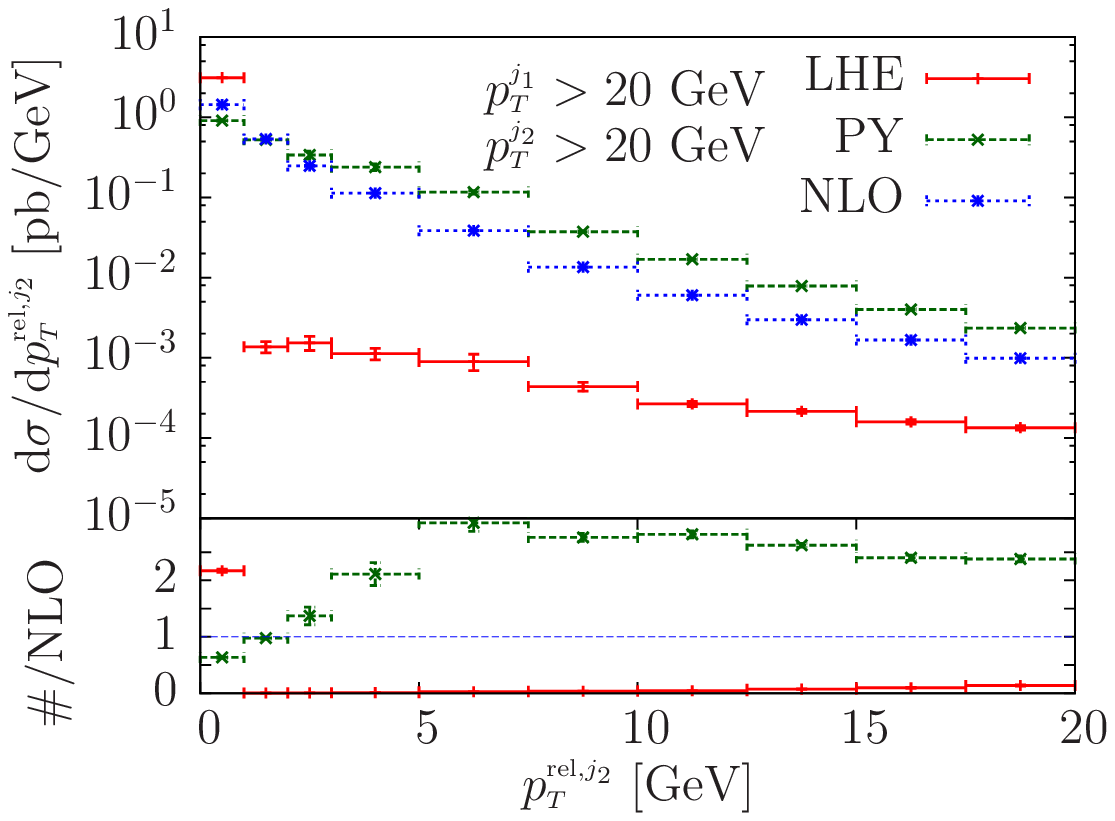}  \nolinebreak
\includegraphics[height=\hfigs]{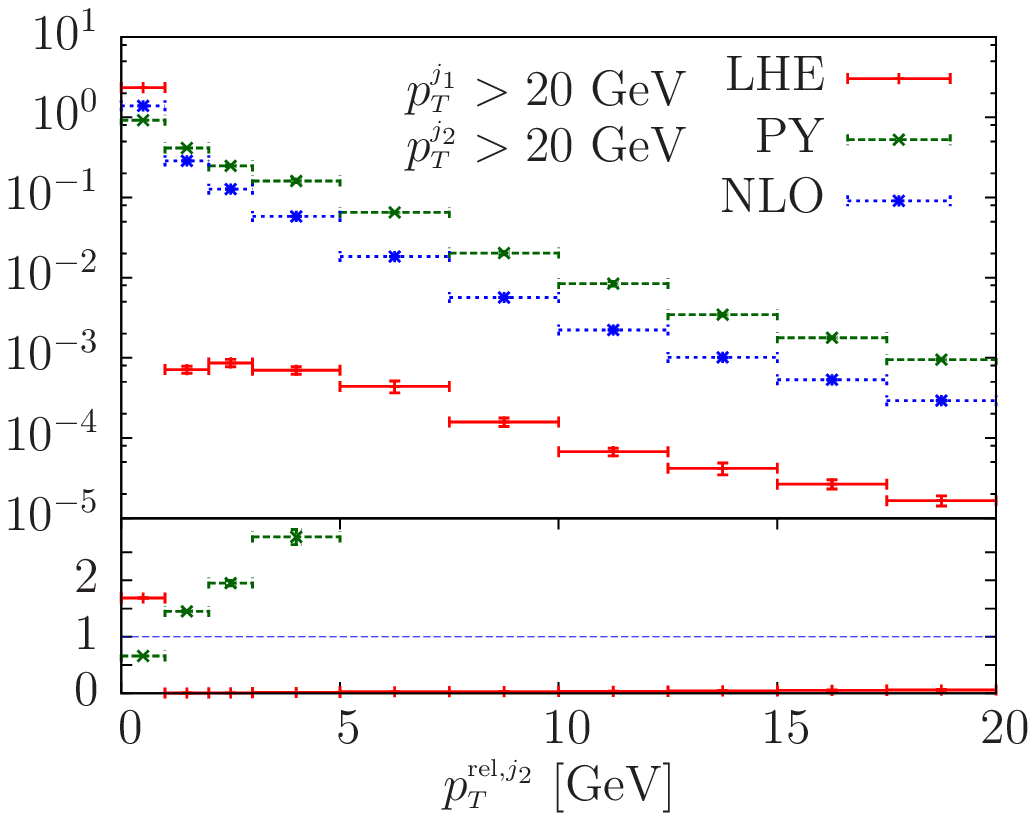} 
\caption{$\ptreltwo$ distribution of the second hardest jet in the $Hjj$
  process. The left and right plots use respectively the $\mH$ and $\HThat$
  scales.}
\label{fig:HJJ_ptrel2}
\end{center}
\end{figure}
In figs.~\ref{fig:HJJ_ptrel1} and~\ref{fig:HJJ_ptrel2} we show the $\ptrel$
distribution for the hardest and second hardest jet respectively. Here too
the pattern is similar to what already observed for the $Hj$ case. The LHE
distribution is strongly suppressed, due to the fact that small $\ptrel$
values also imply that no initial-state radiation has taken place above that
scale, and the distribution is dominated by the shower effects, that are not
obtained with the same scale choice as the NLO result. We note again that the
$\bar{B}/B$ $K$-factor plays a role here, especially for the right plots,
that have a larger value because of the use of the $\HThat$ scale.

\subsection{Hadronization effects and matching ambiguities}
\begin{figure}[htb]
\begin{center}
\includegraphics[height=\hfigs]{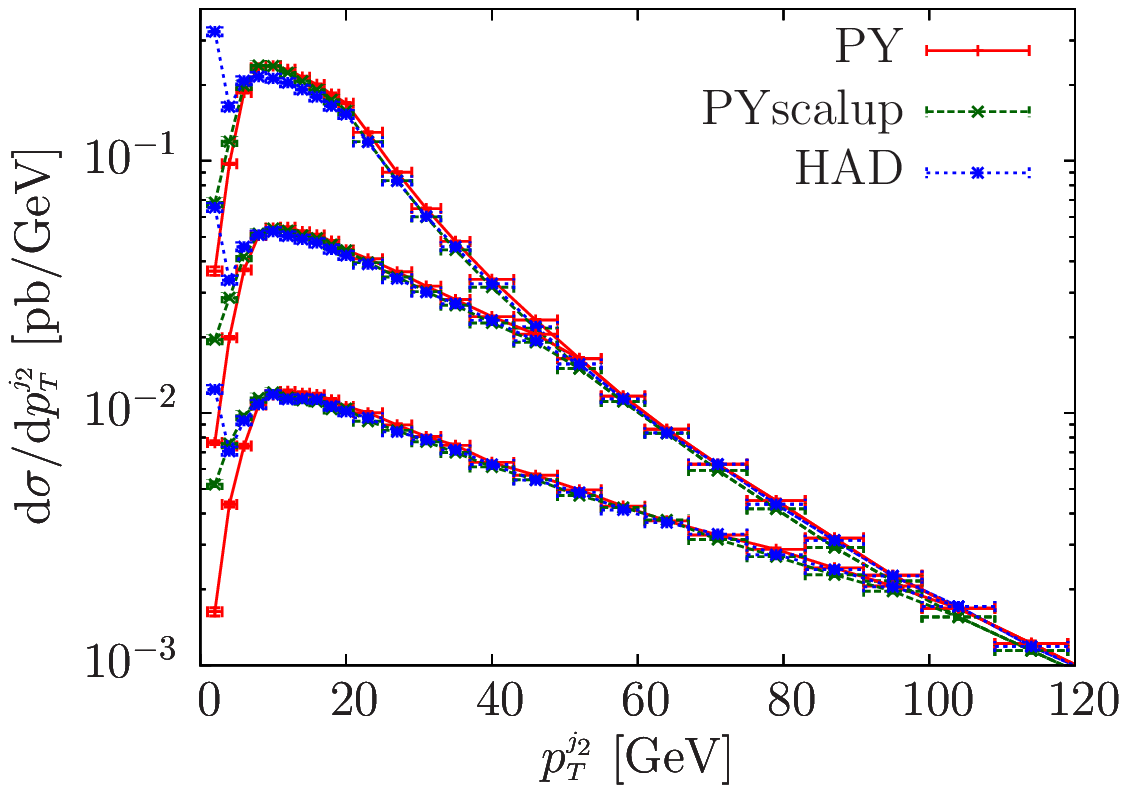} \nolinebreak
\includegraphics[height=\hfigs]{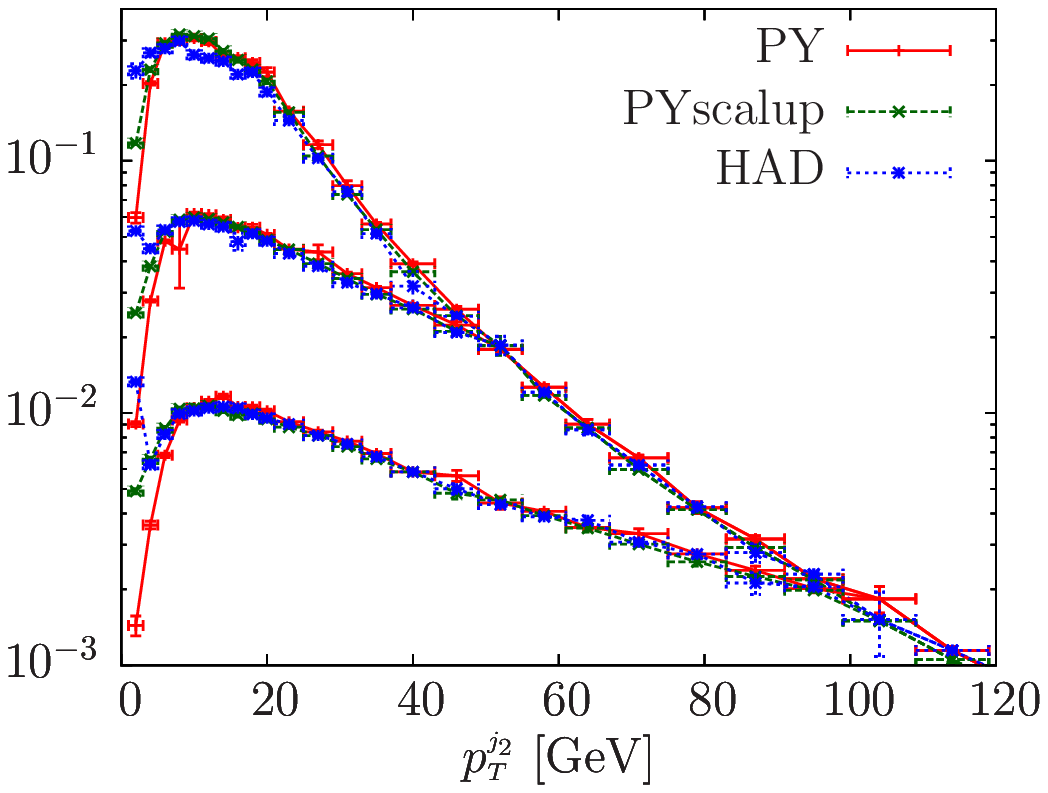}
\caption{Transverse-momentum distribution of the second hardest jet in the
  $Hj$ process for the $\pT$ cut on the hardest jet equal to
  $20$, $50$ and $100$~GeV.}
\label{fig:HJ_j2-pt_cuts_had}
\end{center}
\end{figure}

\begin{figure}[htb]
\begin{center}
\includegraphics[height=\hfigs]{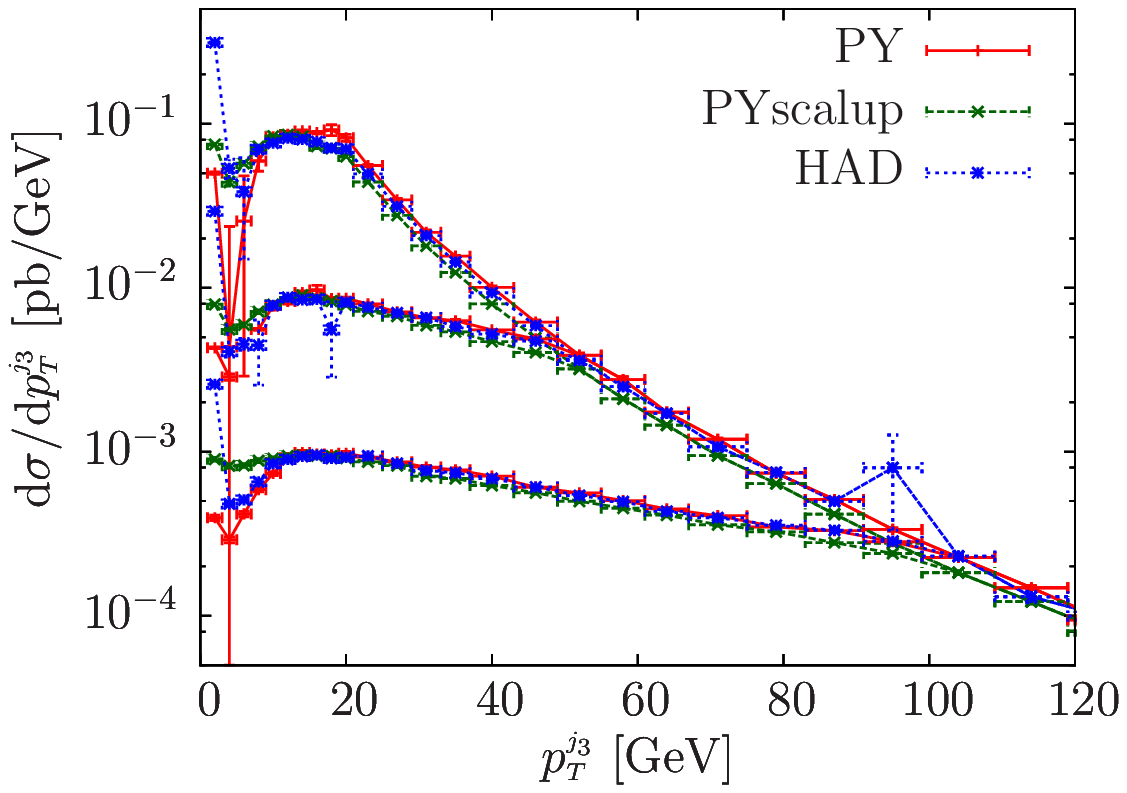}  \nolinebreak
\includegraphics[height=\hfigs]{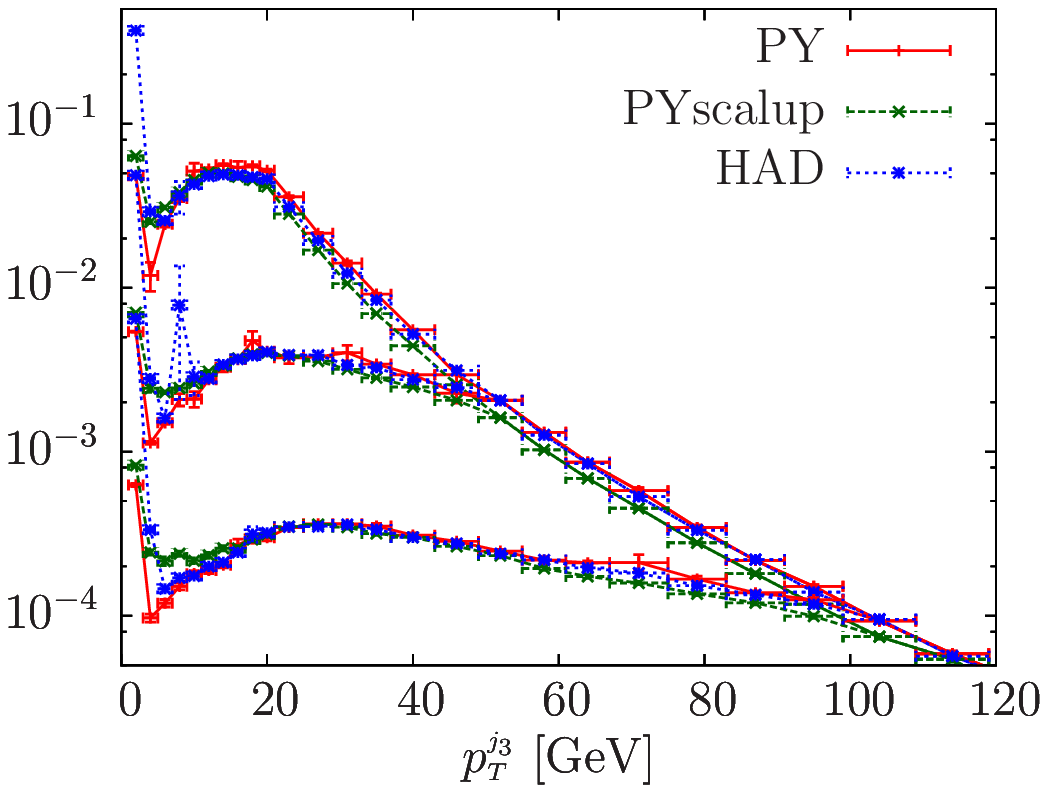} 
\caption{Transverse-momentum distribution of the third hardest jet in the
  $Hjj$ process for the $\pT$ cut on the two hardest jets equal
  to $20$, $50$ and $100$~GeV.}
\label{fig:HJJ_j3-pt_had_cuts}
\end{center}
\end{figure}

In this section we focus upon two topics: the effects of hadronization and
the ambiguities related to matching the LHE result to the shower.  In
figs.~\ref{fig:HJ_j2-pt_cuts_had} and~\ref{fig:HJJ_j3-pt_had_cuts} we display
the transverse-momentum distribution of the radiated jet for different cuts
on the first jet $\pT$, for the $Hj$ and $Hjj$ generators respectively.

The curves labelled ``PYscalup'' are obtained with a non-default
determination of the {\tt scalup} parameter to be set in the Les Houches
interface, which limits the hardness of the radiation of the following
shower. While normally in \POWHEG{} {\tt scalup} is set to the hardest
momentum of the radiation that \POWHEG{} generates, in the ``PYscalup''
results, we determine it by finding the smallest transverse momentum of the
LHE, which can be either the transverse momentum of any parton relative to
the beam axis, or of any parton relative to any other parton, computed in the
center-of-mass frame. Because of the way the real-radiation contribution is
separated into singular contributions, the two choices differ only for
subleading configurations, and one expects only a minor effect due to this
change.  The plots in both figures confirm this expectation. Similarly, we
see that hadronization and underlying-event effects (indicated with HAD in
the figures) have a sizable impact on the distributions only for small
transverse momenta, as expected.

\subsection{Comparison between the $H$, $Hj$ and $Hjj$ generators}
\label{sec:comparison}
In this section, we compare a few distributions that are described by more
than one available \POWHEGBOX{} generator. This comparison can be considered
as a first step in the direction of merging \POWHEGBOX{} samples with
increasing number of jets.


\begin{figure}[htb]
\begin{center}
\includegraphics[height=0.5\textwidth]{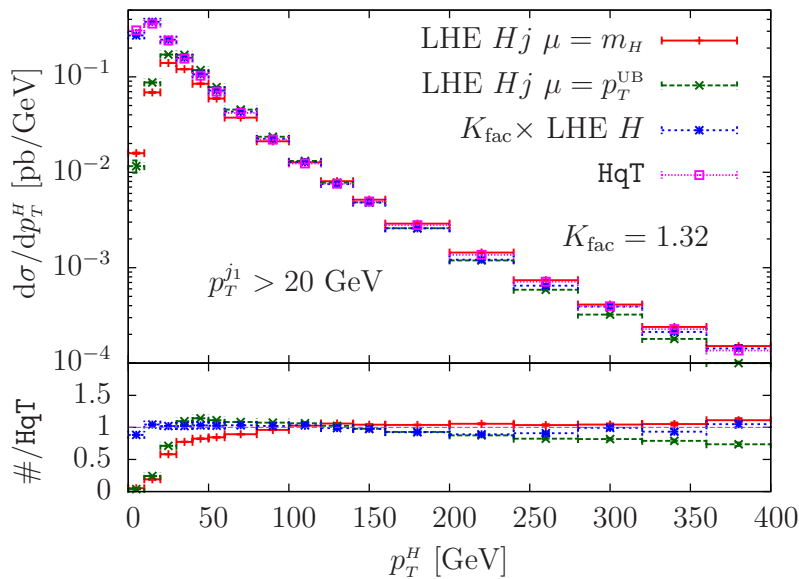} 
\caption{Comparison of the Higgs boson transverse-momentum distribution
  computed with the $H$, $Hj$ and \HqT{} generators. The LHE $Hj$ results are
  shown for the $\mH$ and the Higgs underlying-Born $\pT^{\sss\rm UB}$
  scale choice.}
\label{fig:H-pt_comp}
\end{center}
\end{figure}
We first consider a comparison between the Higgs boson transverse-momentum
spectrum obtained using the $H$ and the $Hj$ generators. The $H$ generator
describes this distribution in the whole $\pT$ range, with the Sudakov region
obtained with next-to-leading logarithmic accuracy, and with leading-order
accuracy for the high-momentum region.  On the other hand, the $Hj$ generator
describes this distribution with NLO accuracy, but only for transverse
momenta above the Sudakov region.

In these comparisons, the $H$ sample was obtained following the
recommendation of ref.~\cite{Dittmaier:2012vm}, i.e.~the parameter {\tt
  hfact} was set to $\mH/1.2$ in the {\tt powheg.input} file. With this
setup, the resulting Higgs boson $\pT$ distribution is in remarkable
agreement with the output of the \HqT{} program~\cite{hqt, Bozzi:2003jy,
  Bozzi:2005wk, deFlorian:2011xf}.  In addition, we apply a $K$-factor of
$1.32$ to the $H$ result, in order to match the \HqT{} total NNLO cross
section ($\sigma_{\sss H}^{\sss \rm NLO} = 10.85$~pb, $\sigma_{\sss H}^{\sss
  \rm NNLO} = 14.35$~pb).  The results are displayed in
fig.~\ref{fig:H-pt_comp}, which demonstrates the agreement between the
results for the $H$ sample and the output of the \HqT{} program, for the
Higgs boson $\pT$ distribution~\cite{Dittmaier:2012vm}. This comparison is
performed using a scale equal to the Higgs boson mass, since \HqT{} accepts
only a fixed renormalization and factorization scale.  We also show two
predictions for the $Hj$ generator, using both the fixed and running-scale
choices.  The $Hj$ generator predictions with a fixed scale are also in
excellent agreement with \HqT{} for large transverse momenta. This is not
surprising, since, in this kinematic region, it is using the same ${\cal
  O}(\as^4)$ result as \HqT{}.  At low $\pT$, the $Hj$ results begin to feel
the lack of soft-gluon resummation effects that are included in \HqT{}.
The $Hj$ result computed with the dynamical scale shows a substantially
similar pattern, undershooting the \HqT{} one by about $25$\%. 

\begin{figure}[htb]
\begin{center}
\includegraphics[height=\hfigs]{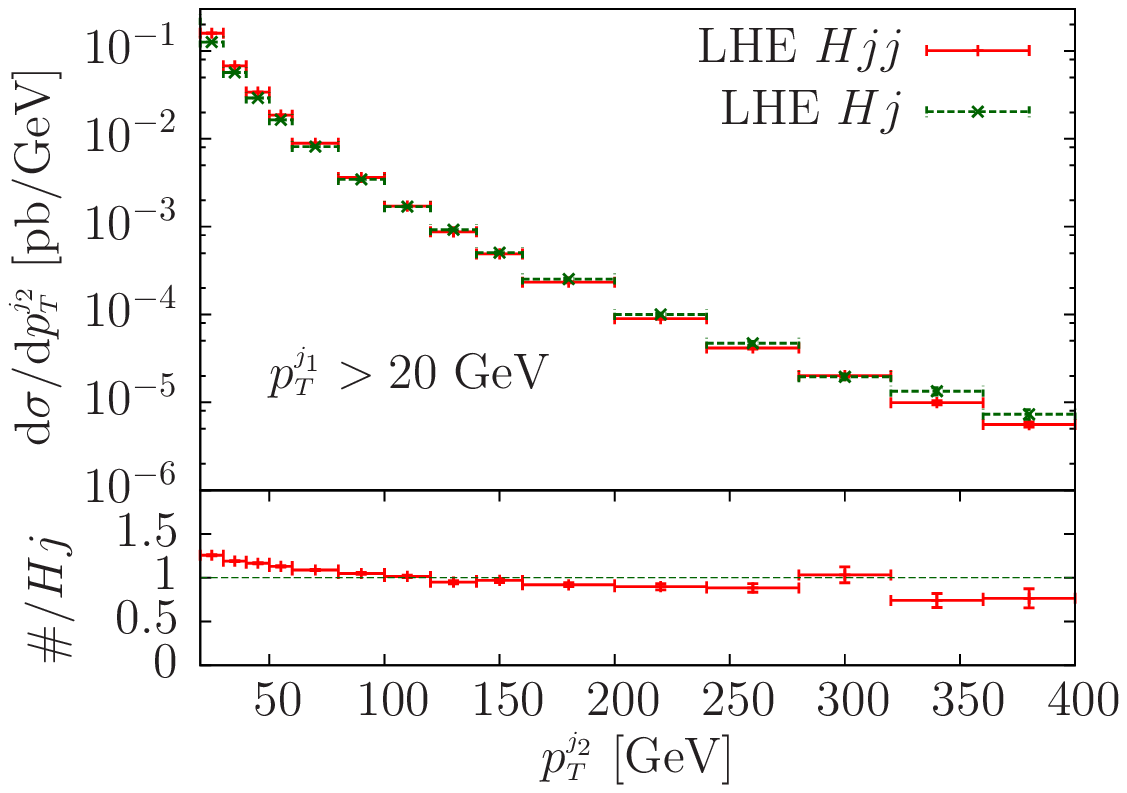} \nolinebreak
\includegraphics[height=\hfigs]{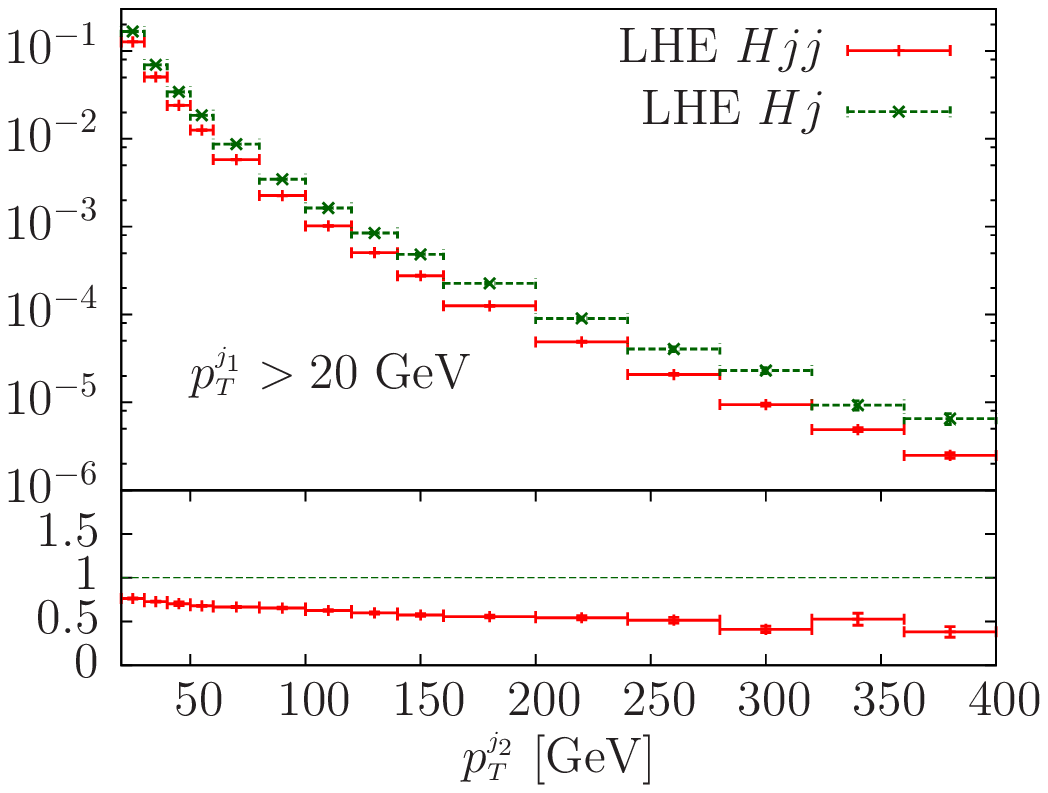} 
\caption{Comparison of the second hardest jet transverse momentum computed
  with the $Hj$ and $Hjj$ generators. The left plot uses the $\mH$ scale in
  both the $Hj$ and $Hjj$ generators. In the right plot the $Hj$ generator
  uses the Higgs underlying-Born transverse-momentum scale, while the $Hjj$
  generator uses $\HThat$.}
\label{fig:j2-pt_comp}
\end{center}
\end{figure}
Turning to the comparison of the $Hj$ and $Hjj$ generators for the
transverse-momentum distribution of the second hardest jet, we notice that
the $Hj$ generator computes this distribution down to small values of the
transverse momentum, since it includes the appropriate Sudakov effects.  At
large transverse momenta, however, it only has leading-order accuracy.  The
$Hjj$ program has instead NLO accuracy at large transverse momenta, but does
not include fully resummed Sudakov effects at small $\pT$.  We do not have
any higher-accuracy calculation for this distribution, as in the previous
case. The results are displayed in fig.~\ref{fig:j2-pt_comp}.  Notice that,
for the $\mH$ scale choice, the matching is very good, something that we
expect since the $K$-factor is close to one in this case. Matching with the
running scales leads, instead, to non-negligible differences, although a
stability plateau is present roughly above $60$~GeV, provided that the
transverse momentum is not too large.

\section{Conclusions}
\label{sec:conclusions}
In the present work we have developed generators for the gluon-fusion
production of a Higgs boson in association with one or two jets, in the large
top-quark mass limit. We have examined the output of the two generators,
comparing them with fixed NLO calculations, among each other, and with the
fully inclusive Higgs boson \POWHEG{} generator.  The features of the
distributions we have examined are all well understood and reflect our
expectations for a typical \POWHEG{} generator. In general, we find that
quantities inclusive in the radiated jet (i.e.~the second jet in $H+1$~jet
and the third jet in $H+2$~jets) are in good agreement with the NLO result,
while distributions in the radiated jet reflect the features of Sudakov
suppression and NLO enhancement that are typically found in NLO generators
matched with a shower. We see no indication of problems related to the
increase in complexity when going from the inclusive Higgs boson production
to the associated production with one and two jets, other than an increase in
the amount of computer time required for the calculation.

The development of the Higgs boson production code has been achieved using a
new interface to the \MG{} code, that has also been presented in this
work. The use of this interface has considerably simplified the construction
of the generator. The interface is fully generic, and so we expect that the
implementation of new processes will be greatly simplified with its use. In
addition, the code takes advantage of compact expressions for Higgs boson
plus four parton virtual amplitudes that had previously been collected in
\MCFM{}.

The code of our new Higgs boson production generators can be accessed via the
\POWHEGBOX{} svn repository (see the \POWHEGBOX{} web page
\url{http://powhegbox.mib.infn.it} for instructions). The new \MG{} interface
is also available there, so that people willing to develop their own
\POWHEG{} code may benefit from its use.


\appendix
\section{The interface with \MG: technical details}
\label{app:tech_details}
The \MG{} interface is available under the {\tt POWHEG-BOX/\MGS} directory.
To use the interface, create a process directory in the {\tt POWHEG-BOX}
directory, and copy the whole content of the \MGS\ directory to this new
process directory. To specify the process, edit the file
\verb|Cards/proc_card.dat| and set the process and the physics model as in
\MG. Always enter the real emission process, i.e.~the Born process plus an
extra jet \verb|j|. For example, to generate Higgs boson plus two jets at a
proton-proton collider, we entered
\begin{verbatim}
   p p > H j j j
   QCD=3
   QED=0
   HIG=1
\end{verbatim}
It is recommended to set the parameters \verb|QCD| and \verb|QED| to be
exactly equal to the number of strong and electroweak interactions in the
real-emission process (excluding the couplings present in the effective
vertex, if any). In the case of the \verb|heft| model (see below), it is also
needed to set the parameter \verb|HIG=1| to allow for the effective Higgs
boson to gluon coupling to be included in the Feynman diagrams.

The ordering of the particles in the process should follow the \POWHEGBOX{}
conventions: 
\begin{enumerate}
  \item first particle: incoming particle with positive rapidity
  
  \item second particle: incoming particle with negative rapidity
  
  \item from the third particle onward: final-state particles ordered as
  follows
  \begin{itemize}
    \item colourless particles first,
    
    \item massive coloured particles,
    
    \item massless coloured particles.
  \end{itemize}
\end{enumerate}
The default \MG\ interface has been validated to work with the
following three physics models:
\begin{itemize}
\item \verb|sm|: this is the default \MG\ model with a massive top
  and bottom quark, diagonal CKM matrix and massless electrons and
  muons.
\item \verb|smckm|: this is the same as the \verb|sm| model, however,
  the full CKM matrix can be specified in the parameter card.
\item \verb|heft|: this is the same as the default \verb|sm| model with the
  inclusion of the effective coupling between gluons and the Higgs boson in
  the large top-quark mass limit. The Higgs boson Yukawa interaction with
  bottom quarks is neglected in this model.
\end{itemize}
Although the interface has not been tested with other physics models,
it is straightforward to extend the interface to work with any simple
extension of the Standard Model.

To generate the process, execute the \verb|NewProcess.sh| script. This will
compile the \MG\ code, generate the (correlated) Born and real-emission
squared amplitudes and create the libraries that can be linked to the
\POWHEGBOX. In particular, \verb|libmadgraph.a| contains the matrix elements,
\verb|libdhelas3.a| the \HELAS{} routines and \verb|libmodel.a| the physics
model. Furthermore, the following files are written in the current directory:
\begin{itemize}
\item \verb|Born.f|: it contains the routines \verb|setborn| to compute
  the (correlated) Born squared matrix elements and
  \verb|borncolour_lh| that assigns the colour flow to a Born process.
\item \verb|real.f|: it contains the routines \verb|setreal| to compute
  the real-emission squared matrix elements.
\item \verb|Cards/param_card.dat|: this is the input card where all the model
  parameters need to be specified.
\item \verb|init_couplings.f|: this contains the \verb|init_couplings|
  routine that sets all the model parameters specified in the
  \verb|param_card.dat|. The coupling constants that depend on the strong
  coupling \verb|st_alpha| or from the event kinematics 
should be specified in the routine \verb|set_ebe_couplings|, that is updated
event-by-event.
\item \verb|coupl.inc|: it contains the common blocks for all the couplings
  used by \MG.
\end{itemize}
To complete the implementation of a process in the \POWHEGBOX{}, the user
must provide the Born phase-space in the file \verb|Born_phsp.f| and the
virtual squared matrix elements in the file \verb|virtual.f|. Also, no
information on possible intermediate resonances in the matrix elements is
kept. The user needs to specify explicitly in the routine \verb|finalize_lh|
(in the \verb|Born.f| file) which resonances should be written in the LHE
file, so that the shower Monte Carlo program can deal with them correctly.
Finally, the resulting code can be compiled  by executing the command
\begin{verbatim}
   $ make pwhg_main
\end{verbatim}

\section{\PYTHIA{} setup}
\label{app:PY_setup}
The sequence of \PYTHIA{} calls we have used in the calculation of the
results presented in sec.~\ref{sec:phenomenology} is the following:
\begin{itemize}
\item  without hadronization
\begin{verbatim}
         call PYTUNE(320)
         call PYINIT('USER','','',0d0)
c Hadronization off
         mstp(111)=0
c primordial kt off
         mstp(91)=0
c No multiple parton interactions
         if(mstp(81).eq.1) then
c Q2 ordered shower
            mstp(81)=0
         elseif(mstp(81).eq.21) then
c p_T^2 ordered shower
            mstp(81)=20
         endif
         call PYABEG
         call PYEVNT
         call PYANAL
\end{verbatim}

\item with hadronization 
\begin{verbatim}
         call PYTUNE(320)
         call PYINIT('USER','','',0d0)
c switching off MPI
         mstp(81)=20
         call PYABEG
         call PYEVNT
         call PYANAL
\end{verbatim}
\end{itemize}

\bibliography{paper}

\providecommand{\href}[2]{#2}\begingroup\raggedright\begin{thebibliography}{10}

\bibitem{Collaboration:2012si}
A.~Collaboration, {\it {Combined search for the Standard Model Higgs boson
  using up to 4.9 fb$^{-1}$ of pp collision data at $\sqrt{s} = 7$~TeV with the
  ATLAS detector at the LHC}},  \href{http://xxx.lanl.gov/abs/1202.1408}{{\tt
  arXiv:1202.1408}}.

\bibitem{Chatrchyan:2012tx}
{\bf CMS Collaboration} Collaboration, S.~Chatrchyan {\em et.~al.}, {\it
  {Combined results of searches for the standard model Higgs boson in pp
  collisions at $\sqrt{s} = 7$~TeV}},
  \href{http://xxx.lanl.gov/abs/1202.1488}{{\tt arXiv:1202.1488}}.

\bibitem{Zeppenfeld:2000td}
D.~Zeppenfeld, R.~Kinnunen, A.~Nikitenko, and E.~Richter-Was, {\it {Measuring
  Higgs boson couplings at the LHC}},  {\em Phys. Rev.} {\bf D62} (2000)
  013009, [\href{http://xxx.lanl.gov/abs/hep-ph/0002036}{{\tt
  hep-ph/0002036}}].

\bibitem{Plehn:2001nj}
T.~Plehn, D.~L. Rainwater, and D.~Zeppenfeld, {\it {Determining the structure
  of Higgs couplings at the LHC}},  {\em Phys. Rev. Lett.} {\bf 88} (2002)
  051801, [\href{http://xxx.lanl.gov/abs/hep-ph/0105325}{{\tt
  hep-ph/0105325}}].

\bibitem{Duhrssen:2004cv}
M.~Duhrssen, S.~Heinemeyer, H.~Logan, D.~Rainwater, G.~Weiglein, {\em et.~al.},
  {\it {Extracting Higgs boson couplings from CERN LHC data}},  {\em Phys.Rev.}
  {\bf D70} (2004) 113009, [\href{http://xxx.lanl.gov/abs/hep-ph/0406323}{{\tt
  hep-ph/0406323}}].

\bibitem{Figy:2003nv}
T.~Figy, C.~Oleari, and D.~Zeppenfeld, {\it {Next-to-leading order jet
  distributions for Higgs boson production via weak-boson fusion}},  {\em Phys.
  Rev.} {\bf D68} (2003) 073005,
  [\href{http://xxx.lanl.gov/abs/hep-ph/0306109}{{\tt hep-ph/0306109}}].

\bibitem{Ravindran:2002dc}
V.~Ravindran, J.~Smith, and W.~L. Van~Neerven, {\it {Next-to-leading order QCD
  corrections to differential distributions of Higgs boson production in hadron
  hadron collisions}},  {\em Nucl. Phys.} {\bf B634} (2002) 247--290,
  [\href{http://xxx.lanl.gov/abs/hep-ph/0201114}{{\tt hep-ph/0201114}}].

\bibitem{Berger:2004pca}
E.~L. Berger and J.~M. Campbell, {\it {Higgs boson production in weak boson
  fusion at next-to- leading order}},  {\em Phys. Rev.} {\bf D70} (2004)
  073011, [\href{http://xxx.lanl.gov/abs/hep-ph/0403194}{{\tt
  hep-ph/0403194}}].

\bibitem{Klamke:2007cu}
G.~Klamke and D.~Zeppenfeld, {\it {Higgs plus two jet production via gluon
  fusion as a signal at the CERN LHC}},  {\em JHEP} {\bf 04} (2007) 052,
  [\href{http://xxx.lanl.gov/abs/hep-ph/0703202}{{\tt hep-ph/0703202}}].

\bibitem{DelDuca:2004wt}
V.~Del~Duca, A.~Frizzo, and F.~Maltoni, {\it {Higgs boson production in
  association with three jets}},  {\em JHEP} {\bf 05} (2004) 064,
  [\href{http://xxx.lanl.gov/abs/hep-ph/0404013}{{\tt hep-ph/0404013}}].

\bibitem{DelDuca:2006hk}
V.~Del~Duca {\em et.~al.}, {\it {Monte Carlo studies of the jet activity in
  Higgs + 2~jet events}},  {\em JHEP} {\bf 10} (2006) 016,
  [\href{http://xxx.lanl.gov/abs/hep-ph/0608158}{{\tt hep-ph/0608158}}].

\bibitem{Campbell:2006xx}
J.~M. Campbell, R.~K. Ellis, and G.~Zanderighi, {\it {Next-to-leading order
  Higgs + 2 jet production via gluon fusion}},  {\em JHEP} {\bf 10} (2006) 028,
  [\href{http://xxx.lanl.gov/abs/hep-ph/0608194}{{\tt hep-ph/0608194}}].

\bibitem{Campbell:2010cz}
J.~M. Campbell, R.~Ellis, and C.~Williams, {\it {Hadronic production of a Higgs
  boson and two jets at next-to-leading order}},  {\em Phys.Rev.} {\bf D81}
  (2010) 074023, [\href{http://xxx.lanl.gov/abs/1001.4495}{{\tt
  arXiv:1001.4495}}].

\bibitem{DelDuca:2001eu}
V.~Del~Duca, W.~Kilgore, C.~Oleari, C.~Schmidt, and D.~Zeppenfeld, {\it {H + 2
  jets via gluon fusion}},  {\em Phys. Rev. Lett.} {\bf 87} (2001) 122001,
  [\href{http://xxx.lanl.gov/abs/hep-ph/0105129}{{\tt hep-ph/0105129}}].

\bibitem{DelDuca:2001fn}
V.~Del~Duca, W.~Kilgore, C.~Oleari, C.~Schmidt, and D.~Zeppenfeld, {\it
  {Gluon-fusion contributions to H + 2 jet production}},  {\em Nucl. Phys.}
  {\bf B616} (2001) 367--399,
  [\href{http://xxx.lanl.gov/abs/hep-ph/0108030}{{\tt hep-ph/0108030}}].

\bibitem{Hoeche:2011fd}
S.~Hoeche, F.~Krauss, M.~Schonherr, and F.~Siegert, {\it {A critical appraisal
  of NLO+PS matching methods}},  \href{http://xxx.lanl.gov/abs/1111.1220}{{\tt
  arXiv:1111.1220}}.

\bibitem{Alioli:2010xd}
S.~Alioli, P.~Nason, C.~Oleari, and E.~Re, {\it {A general framework for
  implementing NLO calculations in shower Monte Carlo programs: the POWHEG
  BOX}},  {\em JHEP} {\bf 06} (2010) 043,
  [\href{http://xxx.lanl.gov/abs/1002.2581}{{\tt arXiv:1002.2581}}].

\bibitem{MCFM}
\texttt{http://mcfm.fnal.gov}.

\bibitem{Stelzer:1994ta}
T.~Stelzer and W.~F. Long, {\it Automatic generation of tree level helicity
  amplitudes},  {\em Comput. Phys. Commun.} {\bf 81} (1994) 357--371,
  [\href{http://xxx.lanl.gov/abs/hep-ph/9401258}{{\tt hep-ph/9401258}}].

\bibitem{Alwall:2007st}
J.~Alwall {\em et.~al.}, {\it {MadGraph/MadEvent v4: The New Web Generation}},
  {\em JHEP} {\bf 09} (2007) 028,
  [\href{http://xxx.lanl.gov/abs/0706.2334}{{\tt arXiv:0706.2334}}].

\bibitem{Murayama:1992gi}
H.~Murayama, I.~Watanabe, and K.~Hagiwara, {\it Helas: Helicity amplitude
  subroutines for feynman diagram evaluations}, . KEK-91-11.

\bibitem{Frederix:2008hu}
R.~Frederix, T.~Gehrmann, and N.~Greiner, {\it {Automation of the Dipole
  Subtraction Method in MadGraph/MadEvent}},  {\em JHEP} {\bf 09} (2008) 122,
  [\href{http://xxx.lanl.gov/abs/0808.2128}{{\tt arXiv:0808.2128}}].

\bibitem{Frederix:2010cj}
R.~Frederix, T.~Gehrmann, and N.~Greiner, {\it {Integrated dipoles with
  MadDipole in the MadGraph framework}},  {\em JHEP} {\bf 1006} (2010) 086,
  [\href{http://xxx.lanl.gov/abs/1004.2905}{{\tt arXiv:1004.2905}}].

\bibitem{Gehrmann:2010ry}
T.~Gehrmann and N.~Greiner, {\it {Photon Radiation with MadDipole}},  {\em
  JHEP} {\bf 1012} (2010) 050, [\href{http://xxx.lanl.gov/abs/1011.0321}{{\tt
  arXiv:1011.0321}}].

\bibitem{Frederix:2009yq}
R.~Frederix, S.~Frixione, F.~Maltoni, and T.~Stelzer, {\it {Automation of
  next-to-leading order computations in QCD: the FKS subtraction}},  {\em JHEP}
  {\bf 10} (2009) 003, [\href{http://xxx.lanl.gov/abs/0908.4272}{{\tt
  arXiv:0908.4272}}].

\bibitem{Wilczek:1977zn}
F.~Wilczek, {\it {Decays of Heavy Vector Mesons Into Higgs Particles}},  {\em
  Phys.Rev.Lett.} {\bf 39} (1977) 1304.

\bibitem{Djouadi:1991tka}
A.~Djouadi, M.~Spira, and P.~Zerwas, {\it {Production of Higgs bosons in proton
  colliders: QCD corrections}},  {\em Phys.Lett.} {\bf B264} (1991) 440--446.

\bibitem{Dawson:1990zj}
S.~Dawson, {\it {Radiative corrections to Higgs boson production}},  {\em
  Nucl.Phys.} {\bf B359} (1991) 283--300.

\bibitem{Berger:2006sh}
C.~F. Berger, V.~Del~Duca, and L.~J. Dixon, {\it {Recursive Construction of
  Higgs-Plus-Multiparton Loop Amplitudes: The Last of the $\Phi$-nite Loop
  Amplitudes}},  {\em Phys.Rev.} {\bf D74} (2006) 094021,
  [\href{http://xxx.lanl.gov/abs/hep-ph/0608180}{{\tt hep-ph/0608180}}].

\bibitem{Badger:2006us}
S.~Badger and E.~Glover, {\it {One-loop helicity amplitudes for H $\to$ gluons:
  The All-minus configuration}},  {\em Nucl.Phys.Proc.Suppl.} {\bf 160} (2006)
  71--75, [\href{http://xxx.lanl.gov/abs/hep-ph/0607139}{{\tt
  hep-ph/0607139}}].

\bibitem{Badger:2007si}
S.~Badger, E.~Glover, and K.~Risager, {\it {One-loop phi-MHV amplitudes using
  the unitarity bootstrap}},  {\em JHEP} {\bf 0707} (2007) 066,
  [\href{http://xxx.lanl.gov/abs/0704.3914}{{\tt arXiv:0704.3914}}].

\bibitem{Glover:2008ffa}
E.~Glover, P.~Mastrolia, and C.~Williams, {\it {One-loop phi-MHV amplitudes
  using the unitarity bootstrap: The General helicity case}},  {\em JHEP} {\bf
  0808} (2008) 017, [\href{http://xxx.lanl.gov/abs/0804.4149}{{\tt
  arXiv:0804.4149}}].

\bibitem{Badger:2009hw}
S.~Badger, E.~Nigel~Glover, P.~Mastrolia, and C.~Williams, {\it {One-loop Higgs
  plus four gluon amplitudes: Full analytic results}},  {\em JHEP} {\bf 1001}
  (2010) 036, [\href{http://xxx.lanl.gov/abs/0909.4475}{{\tt
  arXiv:0909.4475}}].

\bibitem{Dixon:2009uk}
L.~J. Dixon and Y.~Sofianatos, {\it {Analytic one-loop amplitudes for a Higgs
  boson plus four partons}},  {\em JHEP} {\bf 0908} (2009) 058,
  [\href{http://xxx.lanl.gov/abs/0906.0008}{{\tt arXiv:0906.0008}}].

\bibitem{Badger:2009vh}
S.~Badger, J.~M. Campbell, R.~K. Ellis, and C.~Williams, {\it {Analytic results
  for the one-loop NMHV Hqqgg amplitude}},  {\em JHEP} {\bf 0912} (2009) 035,
  [\href{http://xxx.lanl.gov/abs/0910.4481}{{\tt arXiv:0910.4481}}].

\bibitem{Frixione:2007vw}
S.~Frixione, P.~Nason, and C.~Oleari, {\it {Matching NLO QCD computations with
  Parton Shower simulations: the POWHEG method}},  {\em JHEP} {\bf 11} (2007)
  070, [\href{http://xxx.lanl.gov/abs/0709.2092}{{\tt arXiv:0709.2092}}].

\bibitem{Nason:2007vt}
P.~Nason, {\it {MINT: A Computer program for adaptive Monte Carlo integration
  and generation of unweighted distributions}},
  \href{http://xxx.lanl.gov/abs/0709.2085}{{\tt arXiv:0709.2085}}.

\bibitem{Alioli:2010xa}
S.~Alioli, K.~Hamilton, P.~Nason, C.~Oleari, and E.~Re, {\it {Jet pair
  production in POWHEG}},  {\em JHEP} {\bf 04} (2011) 081,
  [\href{http://xxx.lanl.gov/abs/1012.3380}{{\tt arXiv:1012.3380}}].

\bibitem{Pumplin:2002vw}
J.~Pumplin {\em et.~al.}, {\it {New generation of parton distributions with
  uncertainties from global QCD analysis}},  {\em JHEP} {\bf 07} (2002) 012,
  [\href{http://xxx.lanl.gov/abs/hep-ph/0201195}{{\tt hep-ph/0201195}}].

\bibitem{Martin:2009iq}
A.~D. Martin, W.~J. Stirling, R.~S. Thorne, and G.~Watt, {\it {Parton
  distributions for the LHC}},  {\em Eur. Phys. J.} {\bf C63} (2009) 189--285,
  [\href{http://xxx.lanl.gov/abs/0901.0002}{{\tt arXiv:0901.0002}}].

\bibitem{Ball:2008by}
{\bf NNPDF} Collaboration, R.~D. Ball {\em et.~al.}, {\it {A determination of
  parton distributions with faithful uncertainty estimation}},  {\em Nucl.
  Phys.} {\bf B809} (2009) 1--63,
  [\href{http://xxx.lanl.gov/abs/0808.1231}{{\tt arXiv:0808.1231}}].

\bibitem{Cacciari:2008gp}
M.~Cacciari, G.~P. Salam, and G.~Soyez, {\it {The anti-$k_T$ jet clustering
  algorithm}},  {\em JHEP} {\bf 04} (2008) 063,
  [\href{http://xxx.lanl.gov/abs/0802.1189}{{\tt arXiv:0802.1189}}].

\bibitem{Alioli:2008tz}
S.~Alioli, P.~Nason, C.~Oleari, and E.~Re, {\it {NLO Higgs boson production via
  gluon fusion matched with shower in POWHEG}},  {\em JHEP} {\bf 04} (2009)
  002, [\href{http://xxx.lanl.gov/abs/0812.0578}{{\tt arXiv:0812.0578}}].

\bibitem{Nason:2010ap}
P.~Nason, {\it {Recent developments in POWHEG}},  {\em PoS} {\bf RADCOR2009}
  (2010) 018, [\href{http://xxx.lanl.gov/abs/1001.2747}{{\tt
  arXiv:1001.2747}}].

\bibitem{Nason:2012pr}
P.~Nason and B.~Webber, {\it {Next-to-Leading-Order Event Generators}},
  \href{http://xxx.lanl.gov/abs/1202.1251}{{\tt arXiv:1202.1251}}.

\bibitem{Dittmaier:2012vm}
S.~Dittmaier, S.~Dittmaier, C.~Mariotti, G.~Passarino, R.~Tanaka, {\em
  et.~al.}, {\it {Handbook of LHC Higgs Cross Sections: 2. Differential
  Distributions}},  \href{http://xxx.lanl.gov/abs/1201.3084}{{\tt
  arXiv:1201.3084}}. Report of the LHC Higgs Cross Section Working Group.

\bibitem{hqt}
\texttt{http://theory.fi.infn.it/grazzini/codes.html}.

\bibitem{Bozzi:2003jy}
G.~Bozzi, S.~Catani, D.~de~Florian, and M.~Grazzini, {\it {The $q_T$ spectrum
  of the Higgs boson at the LHC in QCD perturbation theory}},  {\em Phys.Lett.}
  {\bf B564} (2003) 65--72, [\href{http://xxx.lanl.gov/abs/hep-ph/0302104}{{\tt
  hep-ph/0302104}}].

\bibitem{Bozzi:2005wk}
G.~Bozzi, S.~Catani, D.~de~Florian, and M.~Grazzini, {\it {Transverse-momentum
  resummation and the spectrum of the Higgs boson at the LHC}},  {\em
  Nucl.Phys.} {\bf B737} (2006) 73--120,
  [\href{http://xxx.lanl.gov/abs/hep-ph/0508068}{{\tt hep-ph/0508068}}].

\bibitem{deFlorian:2011xf}
D.~de~Florian, G.~Ferrera, M.~Grazzini, and D.~Tommasini, {\it
  {Transverse-momentum resummation: Higgs boson production at the Tevatron and
  the LHC}},  {\em JHEP} {\bf 1111} (2011) 064,
  [\href{http://xxx.lanl.gov/abs/1109.2109}{{\tt arXiv:1109.2109}}].

\end{thebibliography}\endgroup

\end{document}